\documentstyle[a4,12pt]{article}
\newcommand{\beq}{\begin{equation}}
\newcommand{\eeq}{\end{equation}}
\newcommand{\beqa}{\begin{eqnarray}}
\newcommand{\eeqa}{\end{eqnarray}}
\newcommand{\no}{\nonumber}

\newcommand{\ul}{\underline}
\newcommand{\ol}{\overline}
\newcommand{\ra}{\rightarrow}

\newcommand{\ve}{\varepsilon}

\newcommand{\dg}{\dagger}

\newcommand{\dfrac}{\displaystyle \frac}

\newcommand{\fsl}{\not\!}

\newcommand{\ben}{\begin{enumerate}}
\newcommand{\een}{\end{enumerate}}
\newcommand{\bfl}{\begin{flushleft}}
\newcommand{\efl}{\end{flushleft}}
\newcommand{\ba}{\begin{array}}
\newcommand{\ea}{\end{array}}
\newcommand{\btab}{\begin{tabular}}
\newcommand{\etab}{\end{tabular}}
\newcommand{\bit}{\begin{itemize}}
\newcommand{\eit}{\end{itemize}}

\normalsize
\begin{document}
\parskip=4pt plus 1pt  
\bibliographystyle{unsrt}
\begin{titlepage}
\begin{flushright}
CERN--TH.6625/92 \\
BUTP--92/38
\end{flushright}

\vspace*{.5cm}
\begin{center} {\Large \bf RADIATIVE SEMILEPTONIC KAON DECAYS}

\vspace{.5cm}
{\bf{J. Bijnens$^{\dagger}$ and G. Ecker$^{\sharp}$}}

\vspace{.2cm}

 CERN \\
 CH$-$1211 Geneva 23 \\

\vspace{.5cm}

  and

\vspace{.5cm}

{\bf{J. Gasser$^{\star}$}}

\vspace{.2cm}

 Institut f\"ur Theoretische Physik\\
Universit\"at Bern, Sidlerstrasse 5 \\
CH$-$3012 Bern\\

\vfill
{\bf Abstract}
\end{center}

\noindent We evaluate the matrix elements for the radiative kaon decays
$K^+ \to l^+\nu_l\gamma$, $l^+ \nu_l l'^+ l'^-$ and
$K \to \pi l \nu_l\gamma$ \ ($l,\ l' = e,\mu$) to next-to-leading order in
chiral perturbation theory.
We calculate  total rates and rates with  several kinematical cuts
and confront the results with existing data.
 Measurements at future
kaon facilities will allow for a more detailed comparison between
theory and experiment.

\vfill
\noindent
{\underline{\hspace{5cm}}}\\

\noindent
$^{\dagger}$ Address after 1 November 1992: NORDITA, Blegdamsvej 17,
DK-2100 Copenhagen,\\
$^{\hspace{1.8ex}}$Denmark.\\
\noindent
$^{\sharp}$ Permanent address:
Institut f\"ur Theoretische Physik, Universit\"at  Wien, Boltz-\\
$^{\hspace{1.8ex}}$manngasse 5,
A--1090 Wien, Austria\\
\noindent
$^{\star}$ Research supported in part by Schweizerischer Nationalfonds\\

\noindent
$^{\hspace{1.8ex}}$CERN-TH.6625/92 \\
$^{\hspace{1.8ex}}$August 1992

\end{titlepage}
\vfill \eject
\pagestyle{empty}
\clearpage\mbox{}\clearpage
\pagestyle{plain}
\setcounter{page}{1}

\section{Introduction}
\label{Intro}
There are two main reasons for undertaking a systematic study of
radiative semileptonic kaon decays at this time:
\bit
\item
Very intense kaon beams will soon become available, which
will increase
the existing, rather meager, data sample dramatically.
In fact, the present investigation was triggered by the experimental
program foreseen at the $\Phi$ factory DAFNE at present under
construction in Frascati \cite{DAFHB}.
\item
The standard model can be tested in those processes to next-to-leading
order in the chiral expansion without any further assumptions. All
low-energy constants involved have already been determined
phenomenologically, allowing for unambiguous predictions for a large
number of available channels. As is the case in general, there is of course
no guarantee that effects of $O(p^6)$ or higher in chiral perturbation
theory (CHPT) \cite{Wein79,GLAP,GLNP1} are small.
However, the full amplitudes of $O(p^4)$ for
radiative semileptonic kaon decays can be directly confronted with
experiment.
\eit

The amplitudes at leading order in CHPT coincide with the current algebra
amplitudes of pre-QCD times. At next-to-leading order,
the amplitudes of $O(p^4)$ carry additional dynamical
information. The one-loop amplitudes entering at $O(p^4)$
are in general divergent and must be supplemented by local
counterterm amplitudes depending on a number of (renormalized) low-energy
coupling constants \cite{GLNP1}. The last part is due to the chiral
anomaly
\cite{ABJB} which contributes to
all the $K$-decay channels considered here. At the mesonic level, all
anomalous amplitudes can be derived from
the Wess--Zumino--Witten functional \cite{WZW} without
any free parameters. Radiative semileptonic decays thus offer a number
of possibilities to
experimentally
investigate
one of the fundamental aspects of the standard model.

The present study completes earlier work by Donoghue and Holstein
\cite{DHH,Hol} who did not calculate the loop amplitudes.
Although non-leading in $1/N_C$,
where $N_C$ is the number of colours, the
importance of\ loop contributions depends very much on the specific
process under consideration. In fact,
radiative semileptonic transitions offer a wide range of possibilities
in this respect:
\bit
\item
Except for the renormalization of the kaon-decay constant,
there is no loop amplitude for $K_{l2\gamma}$.
\item
For $K_{l2l'^+l'^-}$, the one-loop contribution is contained in the
electromagnetic form factor of the charged kaon. This form factor is
known to be dominated by the coupling constant $L_9$ at low energies
\cite{GLNP2}.
\item
The loop amplitude is sizeable for $K_{l3\gamma}$ and interferes
destructively with the local counterterm amplitude. It is essential for
a meaningful comparison with experiment.
\eit

The outline of the paper is as follows. In Sect. \ref{CHPT} we recapitulate
CHPT to next-to-leading order to the extent needed in subsequent
sections. The decay amplitudes for $K_{l2\gamma}$ are calculated in
Sect. 3 to $O(p^4)$ in CHPT. The decays $K_{l2l'^+l'^-}$,
$\pi_{l2e^+e^-}$ and $K_{l3\gamma}$ are treated in Sects. 4 and 5,
respectively. Conclusions are drawn in Sect. 6. Appendix A contains a
summary of our notation and of integrals appearing in the various
one-loop amplitudes. The decomposition of the tensor amplitude for
$K_{l2\gamma}$ into invariant amplitudes is discussed in Appendix B. Finally,
Appendix \ref{ward} contains a derivation of Ward identities without
using formal manipulations of $T$-products and current commutators.
\vspace{1cm}

\renewcommand{\theequation}{\arabic{section}.\arabic{equation}}
\setcounter{equation}{0}

\section{CHPT to next-to-leading order}
\label{CHPT}
Chiral perturbation theory
is a systematic approach to formulate the standard model as a
quantum field theory at the hadronic level.
It is characterized by an effective chiral
Lagrangian in terms of pseudoscalar meson fields (and possibly other
low-lying hadronic states) giving rise to a systematic low-energy
expansion of amplitudes \cite{Wein79,GLAP,GLNP1}.

In the formulation of Ref.~\cite{GLNP1}, one considers the generating
functional $Z[v,a,s,p]$ of
connected Green functions of quark currents associated
with the fundamental Lagrangian
\beq
{\cal L} = {\cal L}^0_{QCD} + \bar q \gamma^\mu (v_\mu + \gamma_5 a_\mu)q
- \bar q (s - i \gamma_5 p)q\ . \label{eq:QCD}
\eeq
Here,
${\cal L}^0_{QCD}$ is the QCD Lagrangian with the masses of the
three light quarks set to zero. The external fields $v_\mu$, $a_\mu$,
$s$ and $p$ are Hermitian $3 \times 3$ matrices in flavour space.
To describe electromagnetic and semileptonic interactions, the
relevant external gauge fields of the standard model are
\footnote{We adopt the present conventions of the Particle Data
Group \cite{PDG}.}
\beqa
r_\mu  =  v_\mu + a_\mu & = & - eQ  A_\mu
\label{eq:gf} \\*
l_\mu  =  v_\mu - a_\mu & = &  - eQ  A_\mu
        - \dfrac{e}{\sqrt{2}\sin{\theta_W}} (W^+_\mu T_+ + \mathrm{h.c.})
        \no
\eeqa $$
Q = \dfrac{1}{3} {\mbox{diag}}(2,-1,-1), \qquad
T_+ = \left( \ba{ccc}
0 & V_{ud} & V_{us} \\
0 & 0 & 0 \\
0 & 0 & 0 \ea \right) \ ,$$
where the $V_{ij}$ are Kobayashi--Maskawa matrix elements.
The quark-mass matrix
\beq
{\cal M} = \mbox{diag}(m_u,m_d,m_s)
\eeq
is contained in the scalar field $s(x)$.

The generating functional $Z$ admits an expansion in powers of
external momenta and quark masses (CHPT).
In the meson sector at leading
order in CHPT, it is given by the classical action
\beq
Z_2 = \int d^4 x {\cal L}_2 (U,v,a,s,p)\ , \eeq
with
${\cal L}_2$    the non-linear $\sigma$ model Lagrangian coupled to the
external fields $v,a,s,p$:
\beq
{\cal L}_2 = \frac{F^2}{4} \langle D_\mu U D^\mu U^\dg +
             \chi U^\dg + \chi^\dg U \rangle \ ,
             \label{eq:L2} \eeq
where
\beq
D_\mu U = \partial_\mu U - ir_\mu U + iU l_\mu, \qquad
\chi = 2 B_0(s + ip), \eeq
and $\langle A \rangle$ stands for the trace of the matrix $A$; $U$ is a
unitary $3 \times 3$ matrix
$$
U^\dg U = {\bf 1}, \qquad \det U = 1,
$$
which
 incorporates the fields of the eight pseudoscalar Goldstone
bosons. A convenient parametrization of it is\footnote{We follow the
Condon--Shortley--de Swart phase conventions.}
\beq
U = \exp{(i\sqrt{2}\Phi/F)},\qquad
\Phi = \left( \ba{ccc}
\dfrac{\pi^0}{\sqrt{2}} + \dfrac{\eta_8}{\sqrt{6}} & -\pi^+ & - K^+ \\*
\pi^- & -\dfrac{\pi^0}{\sqrt{2}} + \dfrac{\eta_8}{\sqrt{6}} & - K^0 \\*
K^- & -\ol{K^0} & - \dfrac{2 \eta_8}{\sqrt{6}} \ea \right). \eeq

The functional $Z_2$ is invariant under local $SU(3)_L\times SU(3)_R$
transformations:
 \beqa
Z_2(v',a',s',p') &=& Z_2(v,a,s,p) \; ,
\label{eq:invariance}
\eeqa
with
\beqa
v'_\mu + a'_\mu &=& g_ R
                    (v_\mu + a_\mu )g^+_R
 + ig_R  \partial_\mu g^+_
R \; \; , \nonumber \\
v'_\mu - a'_\mu &=& g_ L
                    (v_\mu - a_\mu )g^+_L
 + ig_L  \partial_\mu g^+_
L\; \; , \nonumber \\
s'+i p' &=& g_R (s+i p)g^+_L \; \; .
\label{eq:gauge}
\eeqa
The matrix $U$ transforms as $U \rightarrow g_R U
g^+_L$ under (\ref{eq:gauge}) .

   The parameters $F$ and $B_0$ are the only
free constants at $O(p^2)$: $F$ is the pion decay constant
in the chiral limit [cf. Eq. (\ref{eq:FK})] :
\beq
F_\pi =  F[1 + O(m_{quark})] = 93.2 \ \mathrm{MeV}\ , \eeq
whereas $B_0$ is related to the quark condensate:
\beq
\langle 0|\bar u u |0\rangle = - F^2 B_0[1 + O(m_{quark})]\;; \eeq
$B_0$ always appears multiplied by quark masses. At $O(p^2)$, the
product $B_0 m_q$ can be expressed in terms of meson masses because
\beq
M^2_{\pi^+} = B_0 (m_u + m_d) + O({\cal{M}}^2)~. \eeq

The Lagrangian (\ref{eq:L2}) is referred to as the effective chiral
Lagrangian of $O(p^2)$. The chiral counting rules are the following:
the field $U$ is of $O(p^0)$, the derivative $\partial_\mu$ and the
external gauge fields $v_\mu, a_\mu$ are terms of $O(p)$, and the
fields $s, p$ count as $O(p^2)$.

At order $p^4$ the generating functional consists of three different classes of
contributions \cite{GLNP1} :
\ben
\item[i)] The  one-loop diagrams generated by the lowest-order
Lagrangian (\ref{eq:L2}).
\item[ii)] An explicit local action of order $p^4$.
\item[iii)] A contribution to account for the chiral anomaly.
\end{enumerate}

The contributions from categories (i) and (ii) are invariant under
(\ref{eq:gauge}), whereas those from (iii) are not.
Below we will use the Bardeen form \cite{ABJB} of this non-invariant
part. In
this convention, the generating functional is still invariant under
local gauge
transformations generated by the vector currents,
\beq
Z(v',a',s',p') = Z(v,a,s,p) \; \; ; \; g_R=g_L \; .
\label{eq:invtotal}
\eeq

 The local
action of $O(p^4)$ [class (ii)]
is generated by the Lagrangian ${\cal L}_4$
\cite{GLNP1}: \beqa
{\cal L}_4 & = & \dots +~L_4 \langle D_\mu U^\dg D^\mu U \rangle
\langle \chi^\dg U + \chi U^\dg \rangle
 +~L_5 \langle D_\mu U^\dg D^\mu U(\chi^\dg U + U^\dg \chi)
 \rangle  \nonumber \\
& & +\cdots -~i L_9 \langle F_R^{\mu\nu} D_\mu U D_\nu U^\dg +
      F_L^{\mu\nu} D_\mu U^\dg D_\nu U \rangle
+~ L_{10} \langle U^\dg F_R^{\mu\nu} U F_{L\mu\nu}\rangle\ ,
\nonumber \\
\label{eq:L4}
\eeqa
where
\beqa
  F_R^{\mu\nu} & = & \partial^\mu r^\nu -
                     \partial^\nu r^\mu -
                     i [r^\mu,r^\nu]   \\*
  F_L^{\mu\nu} & = & \partial^\mu l^\nu -
                     \partial^\nu l^\mu -
                     i [l^\mu,l^\nu]   \no
\eeqa
and where we have only written down the terms necessary for our purposes.
The low-energy couplings $L_4,L_5, L_9, L_{10}$ are
divergent. They absorb the divergences of the one-loop
graphs leading to amplitudes depending on
renormalized, finite couplings $L_i^r(\mu)$ with the following scale
dependence \beq
L_i^r(\mu_2) =L_i^r(\mu_1) + \frac{\Gamma_i}{16\pi^2} \ln\frac{\mu_1}
{\mu_2} \;\; . \label{eq:scale}
\eeq
Observable quantities are independent of the scale $\mu$, once the loop
contributions are included. The coefficients $\Gamma_i$ are displayed in
Table \ref{tab:Li} together with the phenomenological values of the
renormalized coupling constants $L^r_i(M_\rho)$. In fact, the constants
$L_4$ and $L_5$ will never appear explicitly in our amplitudes because
we will use the relations \cite{GLNP1}
\beqa
\dfrac{F_\pi}{F} & = & 1 - 2 \mu_\pi - \mu_K + \dfrac{4 M^2_\pi}{F^2}
L^r_5(\mu) + \dfrac{4(M^2_\pi + 2 M^2_K)}{F^2} L^r_4(\mu) \no \\*
\dfrac{F_K}{F} & = & 1 - \dfrac{3}{4}\mu_\pi - \dfrac{3}{2}\mu_K
-\dfrac{3}{4}\mu_\eta + \dfrac{4 M^2_K}{F^2} L^r_5(\mu)
+ \dfrac{4(M^2_\pi + 2 M^2_K)}{F^2} L^r_4(\mu) \label{eq:FK}  \no \\*
\mu_P & = & \dfrac{M^2_P}{16 \pi^2 F^2} \log{\dfrac{M_P}{\mu}} \enskip
, \enskip \dfrac{F_K}{F_\pi} \enskip = \enskip 1.22~
\eeqa
which are valid at order $p^4$ in the isospin symmetry limit $m_u=m_d$.

\begin{table}[t]
\begin{center}
\caption{Phenomenological values \protect\cite{GLNP1} for the coupling
constants $L^r_i(M_\rho)$ $(i=4,5,9,10)$. The quantities $\Gamma_i$
determine the scale dependence of the $L^r_i(\mu)$
according to Eq. (\protect\ref{eq:scale}).} \label{tab:Li}
\vspace{.5cm}
\begin{tabular}{|c||r|r|}  \hline
$i$ & $L^r_i(M_\rho) \times 10^3$  & $\Gamma_i$ \\ \hline
  4  & $-$0.3 $\pm$ 0.5  &  1/8  \\
  5  & 1.4 $\pm$ 0.5  & 3/8  \\
 9  & 6.9 $\pm$ 0.7 & 1/4  \\
 10  & $-$5.5 $\pm$ 0.7&   $-$1/4  \\
\hline
\end{tabular}
\end{center}
\end{table}

We now turn to point (iii) above.
The contributions of the chiral anomaly to CHPT amplitudes of $O(p^4)$
can be derived from the Wess--Zumino--Witten functional \cite{WZW}.
Here, we will
only write down the pieces relevant for the radiative decays under
consideration. These terms can be expressed as two anomalous Lagrangians
of $O(p^4)$
\beqa
{\cal L}_{anom} (W\gamma)&=&-\dfrac{i\alpha}{4\sqrt{2}\pi
\sin{\theta_W}} \ve^{\mu\nu\alpha\beta} W^+_\mu F_{\nu\alpha}
\left\langle T_+ \left\{U^\dg D_\beta U,
Q+\frac{1}{2}U^\dg Q U\right\}\right\rangle
\no \\*
&  & +~\mathrm{h.c.} \label{eq:semi} \\
{\cal L}_{anom} (\Phi^3\gamma) & = & - \dfrac{e}{16 \pi^2} \ve^
{\mu\nu\rho\sigma} A_\sigma \langle Q \left[\partial_\mu U \partial_\nu
 U^\dg \partial_\rho U U^\dg - \partial_\mu U^\dg \partial_\nu U
\partial_\rho U^\dg U \right]\rangle  \no \\*
& = & - \dfrac{i e \sqrt{2}}{4 \pi^2 F^3} \ve^{\mu\nu\rho\sigma}
A_\sigma \langle Q \partial_\mu \Phi \partial_\nu \Phi \partial_\rho
\Phi \rangle + \dots\ , \label{eq:ang}
\eeqa
where $D_\mu U = \partial_\mu U + ie A_\mu [Q,U]$ is the covariant
derivative with respect to electromagnetism only and $F_{\mu\nu}=
\partial_\mu A_\nu - \partial_\nu A_\mu$ is the electromagnetic field
strength tensor.

We are now in a position to calculate the amplitudes for radiative
semileptonic $K$ and $\pi$ decays to $O(p^4)$ in CHPT.
We will actually calculate
the various amplitudes for an external $W$, which then turns into a
lepton and its associated neutrino. For the low energies that are
relevant here,
we can neglect the momentum dependence of the $W$ propagator.
\vspace{1cm}
\setcounter{equation}{0}
\setcounter{subsection}{0}

\section{Radiative $K_{l2}$ decays}

We consider the $K_{l2\gamma}$ decay
\beq
K^+ (p) \rightarrow l^+ (p_l) \nu_l (p_\nu) \gamma (q) \hspace{1cm}
[K_{l2\gamma}]\ , \label{k1}
\eeq
where $l$ stands for $e$ or $\mu$, and $\gamma$ is a real photon with $q^2 =
0$. Processes where the (virtual) photon converts into a $e^+ e^-$ or
$\mu^+\mu^-$ pair are considered in the next section. The $K^-$ mode is
obtained from (\ref{k1}) by charge conjugation.

\subsection{Matrix elements and kinematics}

The matrix element for $K^+\rightarrow l^+ \nu_l \gamma$ has the
structure
\beq
T = -iG_F eV_{us}^\star \epsilon^\star_\mu \left \{ F_K L^\mu - H^{\mu \nu}
l_\nu \right \}
\label{k3}
\eeq
with
\beqa
L^\mu &=& m_l \bar{u}(p_\nu) (1 + \gamma_5) \left ( \frac{2p^\mu}{2pq}
- \frac{2 p^\mu_l + \not \!{q} \gamma^\mu}{2 p_l q} \right ) v (p_l)
\nonumber \\
l^\mu &=& \bar{u} (p_\nu)\gamma^\mu  (1 -\gamma_5) v (p_l)
\nonumber \\
H^{\mu \nu} &=& i V (W^2) \epsilon^{\mu \nu \alpha \beta} q_\alpha p_\beta -
A(W^2) (q W g^{\mu \nu} - W^\mu q^\nu)
\nonumber \\
W^\mu &=& (p-q)^\mu = (p_l + p_\nu)^\mu.
\label{k4}
\eeqa

Here, $\epsilon_\mu$ denotes the polarization vector of the photon with
$q^\mu \epsilon_\mu= 0$, whereas $A$,
$V$ stand for two Lorentz-invariant
amplitudes which occur in the general decomposition of the
tensors

\beq
I^{\mu \nu} =
\int dx e^{iqx+iWy} \left\langle
0 \mid T V^\mu_{em} (x) I^\nu_{4-i5}(y) \mid K^+(p)\right\rangle
\; \;, \; \; I=V,A \; \; .
\label{k5}
\eeq
The form factor $A$  $(V)$ is related to the matrix element of
the axial (vector)
current in (\ref{k5}).
 In Appendix \ref{kl2g} we display the general
decomposition of $A^{\mu \nu}$,
$V^{\mu\nu}$ for $q^2 \neq 0$ and provide also the link with the notation
used by the PDG \cite{PDG} and in \cite{ke22,km21}.

The  term proportional to $L^\mu$ in (\ref{k3}) does not contain
unknown quantities -- it
is determined by the amplitude of the non-radiative
decay $K^+ \rightarrow l^+
\nu_l$. This part of the amplitude is usually referred to as "inner
Bremsstrahlung (IB) contribution", whereas the term proportional to $H^{\mu
\nu}$ is called "structure-dependent (SD) part".

The form factors are analytic functions in the complex $W^2$ plane cut
along the positive real axis. The cut starts at $W^2 = (M_K + 2 M_\pi)^2$ for
$A$ [at $W^2 = (M_K + M_\pi)^2$ for $V$]. In our phase convention, $A$ and
$V$ are real in the physical region of $K_{l 2 \gamma}$ decays,
\beq
m_l^2 \leq W^2 \leq M^2_{K}.
\label{k6}
\eeq

The kinematics of (spin-averaged) $K_{l 2 \gamma}$ decays needs two
variables, for which we choose the conventional quantities
\beq
x = 2 p q/M^2_{K} \hspace{0.2cm} , \hspace{0.2cm} y = 2 p  p_l/
M_{K}^2 \; \; .
\label{k15}
\eeq
In the $K$ rest frame, the variable $x$ ($y$) is proportional to the photon
(charged lepton) energy,
\beq
x=2 E_\gamma /M_K \; \; , \; \; y=2 E_l/M_K \; \; ,
\label{k15a}
\eeq
and the angle $\theta_{l\gamma}$ between the photon and the charged lepton is
related to $x$ and $y$ by
\beq
x=\frac{ (1-y/2+A/2)(1-y/2-A/2)}{1-y/2+A/2 {\mbox{cos}} \theta_{l \gamma}} \;
\; ;\  A=\sqrt{y^2-4 r_l} \; \; .
\label{k15b}
\eeq
In terms of these quantities, one has
\beq
W^2 = M^2_{K} (1-x) \hspace{0.2cm} ; \hspace{0.2cm} (q^2 = 0) \hspace{0.2cm} .
\label{k16}
\eeq

We write the physical  region for $x$ and $y$ as
$$
2 \sqrt{r_l} \leq y \leq 1 + r_l
$$
\beq
1 - \frac{1}{2} (y + A) \leq x \leq 1 - \frac{1}{2} (y - A)
\label{k17}
\eeq
or, equivalently, as
\beqa
0 \leq & x &\leq 1-r_l \nonumber \\
1-x +\frac{r_l}{(1-x)} \leq & y & \leq 1+r_l
\label{k17a}
\eeqa
where
\beq
r_l = m^2_l/ M^2_{K} = \left \{ \begin{array}{ll}
1.1 \cdot 10^{-6} (l = e) \\
4.6 \cdot 10^{-2} (l = \mu) \; \; .
\end{array} \right .
\label{k18}
\eeq

  \subsection{Decay rates}

The partial decay rate is
\beq
d\Gamma = \frac{1}{2M_K (2\pi)^5} \sum_{spins} |T|^2
d_{LIPS}(p;p_l,p_\nu,q)\ .
\eeq
The Dalitz plot density
\beq
\rho(x,y) = \frac{d^2\Gamma}{dx dy} = \frac{M_K}{256\pi^3} \sum_{spins} |T|^2
\eeq
is a Lorentz-invariant function which contains $V$ and $A$ in the
following
form \cite{brym},
\beqa
\rho(x,y)&=&
\rho_{\mbox{\tiny{IB}}}(x,y)  +  \rho_{\mbox{\tiny{SD}}}(x,y)
+  \rho_{\mbox{\tiny{INT}}}(x,y)
\nonumber \\
 \rho_{\mbox{\tiny{IB}}}(x,y)& =& A_{\mbox{\tiny{IB}}}
f_{\mbox{\tiny{IB}}}(x,y)
\nonumber \\
 \rho_{\mbox{\tiny{SD}}}(x,y)& =& A_{\mbox{\tiny{SD}}}
      M_K^2   \left[ (V+A)^2
f_{{\mbox{\tiny{SD}}}^+}(x,y) + (V-A)^2 f_{{\mbox{\tiny{SD}}}^-} (x,y) \right]
          \\
 \rho_{\mbox{\tiny{INT}}}(x,y)& =&
 A_{\mbox{\tiny{INT}}} M_K
\left [ (V+A) f_{{\mbox{\tiny{INT}}}^+} (x,y) + (V-A)
f_{{\mbox{\tiny{INT}}}^-} (x,y) \right]\ ,\nonumber
\label{k19}
\eeqa
where
\beqa
f_{\mbox{\tiny{IB}}}(x,y)& =&\left[ \frac{1-y+r_l}{x^2(x+y-1-r_l)}\right]
\left[x^2 +2(1-x)(1-r_l) -\frac{2x r_l (1-r_l)}{x+y-1-r_l} \right]
\nonumber \\
 f_{{\mbox{\tiny{SD}}^+}}(x,y)& =& \left[ x+y-1-r_l\right]
\left[ (x+y-1)(1-x)-r_l \right]
\nonumber \\
 f_{{\mbox{\tiny{SD}}^-}}(x,y)& =& \left[1-y+r_l \right]
 \left[ (1-x) (1-y) +r_l\right]
          \\
 f_{{\mbox{\tiny{INT}}^+}}(x,y)& =& \left[ \frac{1-y+r_l}{x(x+y-1-r_l)}\right]
\left[ (1-x)(1-x-y)+r_l \right]
\nonumber \\
 f_{{\mbox{\tiny{INT}}^-}}(x,y)& =& \left[ \frac{1-y+r_l}{x(x+y-1-r_l)}\right]
\left[ x^2 -(1-x)(1-x-y)-r_l\right]\nonumber
\label{k20}
\eeqa

and
\beqa
A_{\mbox{\tiny{IB}}}& =& {4r_l} \left ( \frac{F_K}{M_K} \right)^2
A_{\mbox{\tiny{SD}}}
\nonumber \\
A_{\mbox{\tiny{SD}}}& =& \frac{G_F^2 |V_{us}|^2 \alpha}{32 \pi^2} M_K^5
          \\
A_{\mbox{\tiny{INT}}}& =&{4r_l} \left ( \frac{F_K}{M_K} \right)
A_{\mbox{\tiny{SD}}} \; \; . \nonumber
\label{k20a}
\eeqa
For later convenience, we note that
\beq
A_{\mbox{\tiny{SD}}}
                    = \frac{\alpha}{8\pi}
\frac{1}{r_l(1-r_l)^2} {\left( \frac{M_K}{F_K} \right) }^2
\Gamma (K\rightarrow l\nu_l) \; .
\label{k20b}
\eeq
The subscripts
IB, SD and INT stand respectively for the contribution from
inner Bremsstrahlung, from the structure-dependent part, and from the
interference term between the IB and the SD parts in the amplitude.

To get a feeling for the magnitude of the various contributions
IB, $\mbox{SD}^\pm$ and ${\mbox{INT}}^\pm$ to the decay rate,
we consider
the integrated  rates
\beq
\Gamma_I = \int_{R_I} dxdy  \rho_I (x,y) \; \; ; \; \; I =
\mbox{SD}^\pm, \mbox{INT}^\pm ,\mbox{IB} \; \; ,
\label{dr1}
\eeq
where $\rho_{SD}=\rho_{SD^+} + \rho_{SD^-}$ etc.
 For the region $R_I$ we take the full phase space
 for $I \neq \mbox{IB}$, and
\beq
R_{{\mbox{{\tiny{IB}}}}} =214.5\  {\mbox{MeV/c}} \leq p_l \leq 231.5
\ {\mbox{MeV/c}}\;  \label{dr2}
\eeq
for the Bremsstrahlung contribution. Here $p_l$ stands for the modulus of the
lepton three-momentum in the kaon rest system \footnote{
This cut has been used in \cite{km21} for $K_{\mu 2\gamma}$, because this
kinematical region is free from $K_{\mu3}$ background.
We apply it here for illustration also to the electron mode $K_{e2\gamma}$.
                                 }.
  We consider constant
form factors
$V$, $A$ and write for the rates and for the corresponding branching ratios
\beqa
\Gamma_I & =& A_{\mbox{\tiny{SD}}} \left \{ M_K (V \pm A) \right \}^{N_I}X_I \;
\;
\nonumber \\
{\mbox{BR}}_I &\doteq& \Gamma_I / \Gamma_{\mbox{\tiny{tot}}} = N \left \{
M_K (V \pm A) \right \}^{N_I} X_I
\label{dr3}
\eeqa
with
\beq
N = A_{\mbox{\tiny{SD}}}/{\Gamma_{\mbox{\tiny{tot}}}}
 = 8.348 \cdot 10^{-2}.
\label{dr3a}
\eeq
The values for $N_I$ and $X_I$ are listed in Table \ref{t:pskl2}.

\begin{table}[t]
\protect
\begin{center}
\caption{
 The quantities
 $X_I,N_I$.
 SD$^\pm$ and INT$^\pm$ are evaluated with full phase space,
 IB with
restricted kinematics (\protect\ref{dr2}).
\label{t:pskl2}         }
\vspace{1em}
\begin{tabular}{|c||c|c|c|c|c||c|} \hline
\multicolumn{1}{|c||}{}&{${\mbox{SD}}^+$}&{${\mbox{SD}}^-$}&
{${\mbox{INT}}^+$}&{INT$^-$}
&{${\mbox{IB}}$}& \multicolumn{1}{|c|}{}\\ \hline
{$X_I$} & {$1.67\cdot 10^{-2}$}&{$1.67\cdot 10^{-2}$}&
{$-8.22 \cdot 10^{-8}$}&{$3.67\cdot 10^{-6}$}&{$3.58\cdot 10^{-6}$}&
{$K_{e2\gamma}$} \\ \hline
{$X_I$} & {$1.18\cdot 10^{-2}$}&{$1.18\cdot 10^{-2}$}&
{$-1.78\cdot 10^{-3}$}&{$1.23\cdot 10^{-2}$}&{$3.68\cdot 10^{-2}$}&
{$K_{\mu 2 \gamma}$} \\ \hline
{$N_I$} &2&2&1&1&0&\multicolumn{1}{|c|}{} \\ \hline
  \end{tabular}
\end{center}
\end{table}

To estimate $\Gamma_I$ and ${\mbox{BR}}_I$, we note that the form factors $V,A$
 are of
 order
\beq
M_K(V+A) \simeq -10^{-1} \; \; , \; \; M_K(V-A) \simeq -4 \cdot 10^{-2} \; \; .
\eeq
{}From this and from the entries in the table
one concludes that for the above regions $R_I$, the interference terms
$\mbox{INT}^\pm$ are negligible in
$K_{e2\gamma}$, whereas they are important in $K_{\mu 2 \gamma}$. Furthermore,
IB is negligible for $K_{e2\gamma}$, because it is helicity-suppressed,
as can
be seen from the factor $m_l^2$ in $A_{\mbox{\tiny{IB}}}$. This term
dominates however in $K_{\mu 2 \gamma}$.

\subsection{Determination of $A(W^2)$ and $V(W^2)$}

The decay rate  contains two real functions
\beq
F^\pm(W^2) = V(W^2) \pm A(W^2)
\label{k21}
\eeq
as the only unknowns.
The density distributions
$f_{\mbox{\tiny{IB}}}, \ldots,$ $f_{{\mbox{\tiny{INT}}^\pm}}$
 have  very different Dalitz plots \cite{brym,BEGRep}. Therefore, in
principle,
one can determine the strength of each term by choosing a suitable kinematical
region of observation. To pin down $F^\pm$, it would   be sufficient to
measure
at each photon energy the interference term INT$^\pm$. This has not yet
been achieved, either because the contribution of INT$^\pm$
is too small (in $K_{e 2 \gamma}$),
  or because too few events have been collected  (in
$K_{\mu 2 \gamma}$). On the other hand, from a
measurement
of {\mbox{SD}}$^\pm$ alone one can determine $A, V$ only up to a fourfold
 ambiguity:
\beq
\mbox{SD}^\pm \rightarrow \left \{ (V,A); - (V,A); (A,V); - (A,V) \right \}.
\label{k22}
\eeq
In terms of the ratio
\beq
\gamma_K = A/V
\label{k23}
\eeq
this ambiguity  amounts to
\beq
\mbox{SD}^\pm \rightarrow \left \{ \gamma_K ; 1/ \gamma_K \right \}.
\label{k24}
\eeq
Therefore, in order to pin down the amplitudes $A$ and $V$ uniquely, one must
measure the interference terms INT$^\pm$ as well.

\subsection{Previous experiments}

 \vspace{.5cm}

{\underline{{\bf{$K^+ \rightarrow e^+ \nu_e  \gamma$}}}}

 \vspace{.5cm}

The PDG uses data from two experiments \cite{ke22,ke21}, both of which have
been
sensitive mainly to the {\mbox{SD}}$^+$ term in (\ref{k19}). In \cite{ke21}, 56
events
with $E_\gamma > 100$ MeV, $E_{e^+} > 236$ MeV and
$\theta_{e^+\gamma}> 120^\circ$
have been identified, whereas the later experiment \cite{ke22} has
collected 51 events with $E_\gamma > 48$ MeV, $E_{e^+} > 235$ MeV, and
$\theta_{e^+ \gamma} > 140^\circ$.
In these kinematical regions, background from
$K^+ \rightarrow e^+ \nu_e \pi^0$ is absent because $E_{e}^{\mbox{\tiny{max}}}
(K_{e3}) = 228$ MeV. The combined result of both experiments is
\footnote{In all four experiments \cite{ke21,ke22,km21,km22} discussed here
and below, the form factors $A$ and $V$ have been treated as constants.}
\cite{ke22}
\beq
\Gamma(\mbox{SD}^+) / \Gamma(K_{\mu2}) = (2.4 \pm 0.36) \cdot 10^{-5}.
\label{k25}
\eeq
For {\mbox{SD}}$^-$, the bound
\beq
\Gamma(\mbox{SD}^-) / \Gamma_{\mbox{\tiny{total}}} < 1.6 \cdot 10^{-4}
\label{k26}
\eeq
has been obtained from a sample of electrons with energies 220 MeV $\leq
E_{e} \leq 230$ MeV \cite{ke22}. Using (\ref{dr3},\ref{dr3a}), the
result (\ref{k25}) leads to
\beq
 M_K \mid V+A \mid \;= 0.105 \pm 0.008 \; \; .
\label{k27}
\eeq

The bound (\ref{k26}) on the other hand implies \cite{ke22}
\beq
\mid V -A \mid / \mid V+A \mid < \sqrt{11},
\label{k28}
\eeq
from where it is concluded \cite{ke22}
that $\gamma_K$ is outside the range
$-1.86$ to $-0.54$,
\beq
\gamma_K  \not \in [-1.86, - 0.54] \; \; .
\label{k29}
\eeq
As we already mentioned, the interference terms INT$^\pm$ in $K \rightarrow
e \nu_e \gamma$ are small and can hardly ever be measured.
As a result of this,
the amplitudes $A,V$ and the ratio $\gamma_K$ determined from $K_{e2\gamma}$
are
subject to the ambiguities (\ref{k22}), (\ref{k24}).

\vspace{.5cm}

{\underline{{\bf{$K^+ \rightarrow \mu^+ \nu_\mu \gamma$}}}}

\vspace{.5cm}

Here, the interference terms INT$^\pm$ are non-negligible
in appropriate regions
of phase space, see \cite{brym,BEGRep}. Therefore, this
decay allows one in principle to
pin down $V$ and $A$. The PDG uses data from two experiments
\cite{km21,km22}. In \cite{km21},
the momentum spectrum of the muon was measured in the region (\ref{dr2}).
 In total $2 \pm 3.44$ {\mbox{SD}}$^+$ events
have been found with 216 MeV/c  $< p_\mu< $  230 MeV/c and $E_\gamma > 100$
 MeV, which leads to
\beq
M_K \mid V+ A \mid < 0.16\; \;.
\label{k30}
\eeq
In order to identify
the effect of the {\mbox{SD}}$^-$ terms, the region 120 MeV/c
$<p_\mu < $~150 MeV/c was searched.
Here, the background from $K_{\mu3}$ decays
was very serious. The authors found 142 $K_{\mu \nu \gamma}$ candidates and
conclude that
\beq
-1.77 < M_K (V-A) < 0.21\;.
\label{k31}
\eeq

The result (\ref{k30}) is consistent with
(\ref{k27}), and the bound  (\ref{k31}) is worse than the
result (\ref{k28}) obtained from   $K_{e 2 \gamma}$. The
branching ratios which follow  \cite{km21} from
(\ref{k30}, \ref{k31}) are displayed in Table
\ref{t:erkl2}, where we also show the $K_{e2\gamma}$ results \cite{ke21,ke22}.
The entry SD$^-$+INT$^-$
for $K_{\mu 2 \gamma}$ is based on additional
 constraints from $K_{e2\gamma}$ \cite{km21}.

\begin{table}[t]
\begin{center}
\protect
\caption{ Measured
 branching ratios $\Gamma (K\rightarrow l \nu_l \gamma)
/\Gamma_{\mbox{{\protect\tiny{total}}}}$. The $K_{e2\gamma}$ data are from
\protect\cite{ke21,ke22}, the $K_{\mu 2\gamma}$ data from
\protect\cite{km21,km22}. The last column corresponds \protect\cite{km21}
 to the cut
(\protect\ref{dr2}).
\label{t:erkl2}         }
\vspace{1em}
{\footnotesize{
\begin{tabular}{|c|c|c|c|c|c|} \hline
&{${\mbox{SD}}^+$}&{${\mbox{SD}}^-$}&{${\mbox{INT}}^+$}&
{${\mbox{SD}}^- + {\mbox{INT}}^-$}&{total}
\\ \hline
{$K_{e2\gamma}$} & {$(1.52 \pm 0.23)\cdot 10^{-5}$}&{$<1.6\cdot 10^{-4}$}&
{}&{}&{ }
\\ \hline
{$K_{\mu 2 \gamma}$} & {$< 3\cdot 10^{-5}$}&{}&
{$<2.7\cdot 10^{-5}$}& {$<2.6\cdot 10^{-4}$}&
{$(3.02\pm 0.10)\cdot 10^{-3}$}
\\
\multicolumn{1}{|c|}{}&{}&{}&{(modulus)}&{(modulus)}&{}
\\ \hline
\end{tabular}
}}
\end{center}
\end{table}

\subsection{Theory}

The amplitudes $A(W^2)$ and $V(W^2)$ have been worked out in the framework of
various approaches such as current algebra, PCAC, resonance exchange,
dispersion relations, \ldots . For a rather detailed review together with an
extensive list of references up to 1976, see \cite{BARDIN}. Here, we
concentrate
on the predictions of $V, A$ in the framework of CHPT.

\vspace{.5cm}

{\bf{A) Chiral expansion to one loop}}

\vspace{.5cm}

At leading order in the low-energy expansion, one has
\footnote{
The relevant Feynman diagrams are displayed and discussed in the next section
in connection with the decays
$K^{\pm} \rightarrow l^{\pm} \nu l'^{+}l'^{-}$
where the photon is virtual.}
\beq
A=V=0
\label{k36}
\eeq
and $F_K$ replaced by $F$ in Eq. (\ref{k3}). The rate is therefore
entirely given by the IB contribution at leading order. The loop effects
manifest themselves only in the replacement $F \ra F_K$ in
Eq. (\ref{k3}). The local terms of order $p^4$ give \cite{DHH}
 \beqa
A &=& -\frac{4}{F} (L_9^{r} + L_{10}^{r})
\nonumber \\
V &=& - \frac{1}{8 \pi^2} \frac{1}{F}
\nonumber \\[2mm]
\gamma_K &=& 32 \pi^2 (L_9^r + L_{10}^r)\; ,
\label{k37}
\eeqa
where $L_9^r$ and $L_{10}^r$ are the renormalized low-energy couplings
evaluated at the scale $\mu$ (the combination $L_9^r + L_{10}^r$ is
scale-independent).
The vector form factor stems from the Wess--Zumino term
\cite{WZW}.

\noindent
{\underline {Remarks}:}

\begin{list}%
{AAA}{
\setlength{\leftmargin}{.5cm}%
}
\item[1.]
 At this order in the low-energy expansion, the form factors $A,V$ do
not
exhibit any $W^2$ dependence. A non-trivial
$W^2$ dependence only occurs at the
next order in the energy expansion (two-loop effect, see  the discussion
below). Note that the available analyses of experimental data of $K \rightarrow
l \nu_l \gamma$ decays \cite{ke21,ke22,km21,km22} use constant form factors
throughout.

\item[2.]
 Once the combination $L_9 + L_{10}$ has been pinned down
from other processes, Eq. (\ref{k37}) allows one to
evaluate
$A,V$ unambiguously at this order in the low-energy expansion. Using $L_9 +
L_{10} = 1.4 \cdot 10^{-3}$  and $F=F_\pi$, one has
\beqa
 M_K (A+V) &=& -0.097
\nonumber \\
M_K(V-A) &=& -0.037
\nonumber \\
\gamma_K &=& 0.45 \; \; .
\label{k38}
\eeqa
The result for the combination $(A+V)$ agrees with (\ref{k27}) within the
errors,
while $\gamma_K$ is consistent with (\ref{k29}).

\item[3.]
At this order in the low-energy expansion, the form factors $A,V$
are identical
to the ones in radiative pion decays $\pi^+\rightarrow \l^+\nu_l\gamma $
\cite{GLAP}, as a result of which one has
\beq
 \gamma_\pi = \gamma_K + O({\cal M})\;,
\eeq
\end{list}

We display in Table \ref{t:chpr} the branching ratios ${\mbox{BR}}_I$
 (\ref{dr3})
 which follow from
the prediction (\ref{k38}). These predictions satisfy of course the
inequalities found from experimental data (see Table \ref{t:erkl2}).

\begin{table}
\protect
\begin{center}
\caption{Chiral
prediction at order $p^4$ for the branching ratios
$\Gamma(K\rightarrow l \nu_l \gamma)/\Gamma_{\mbox{\protect\tiny{total}}}$. The
cut used in the last column is  given in Eq. (\protect\ref{dr2}).
\label{t:chpr}
}
\vspace{1em}
\begin{tabular}{|c|c|c|c|c|c|} \hline
{}&{{\mbox{SD}}$^+$}&{{\mbox{SD}}$^-$}&{{\mbox{INT}}$^+$}&{{\mbox{INT}}$^-$}&{
total} \\ \hline
{$K_{e2\gamma}$} & {$1.30\cdot 10^{-5}$}&{$1.95\cdot 10^{-6}$}&
{$6.64\cdot 10^{-10}$}&{$-1.15\cdot 10^{-8}$}&{$2.34 \cdot 10^{-6}$}
\\ \hline
{$K_{\mu 2\gamma}$} & {$9.24\cdot 10^{-6}$}&{$1.38\cdot 10^{-6}$}&
{$1.44\cdot 10^{-5}$}&{$-3.83\cdot 10^{-5}$}&{$3.08 \cdot 10^{-3}$}
\\ \hline
\end{tabular}
\end{center}
\end{table}

\vspace{.5cm}
{\bf{B) $W^2$ dependence of the form factors}}

\vspace{.5cm}

The chiral prediction gives constant form factors at order $p^4$. Terms of
order $p^6$ have not yet been calculated. They would, however, generate a
non-trivial $W^2$ dependence both in $V$ and $A$. In order to estimate the
magnitude of these corrections, we consider one class of $p^6$ contributions:
terms which are generated
by vector and axial vector resonance exchange with
strangeness \cite{BARDIN,EGPR},
\beq
V(W^2) = \frac{V}{1-W^2/{M_{K^\star}}^2}\; \; ,\; \; A(W^2) =
\frac{A}{1-W^2/{M_{K_1}}^2}\;, \label{k40}
\eeq
where $V,A$ are given in (\ref{k37}). We now examine the effect of the
denominators in (\ref{k40}) in the region $y \geq 0.95, x \geq 0.2$
which has been explored in $K^+ \rightarrow e^+ \nu_e \gamma$ \cite{ke22}.
We put $m_e = 0$ and evaluate the rate
\beq
\frac{dP(x)}{dx} = \frac{N_{\mbox{\tiny{tot}}}}{\Gamma_{\mbox{\tiny{tot}}}}
\int^1_{y=0.95}  \rho_{{\mbox{\tiny{SD}}}^+}(x,y) dy \;,
\label{k41}
\eeq
where $N_{\mbox{\tiny{tot}}}$ denotes the total number of $K^+$ decays
considered, and $\Gamma^{-1}_{\mbox{\tiny{tot}}} = 1.24 \cdot 10^{-8}$
s.

\begin{figure}[t]
\vspace{9cm}
\caption{The rate $dP(x)/dx$ in (\protect\ref{k41}), evaluated with the form
factors
(\protect\ref{k40}) and
$N_{\mbox{\protect\tiny{tot}}} = 9
\cdot 10^9$.
The solid line corresponds to $M_{K^\star}
= 890$ MeV, $M_{K_1} = 1.3$ GeV. The dashed  line is evaluated with
$M_{K^\star} = 890$ MeV, $ M_{K_1} = \infty$,
and the dotted line corresponds to
$M_{K^\star} = M_{K_1} = \infty$.
 The total number of events is also indicated in each case.
 \label{figform}
        }
\end{figure}

The function $\frac{dP(x)}{dx}$ is displayed in Fig.  \ref{figform}  for
three different values of $M_{K^\star}$ and $M_{K_1}$, with
 $N_{\mbox{\protect\tiny{tot}}} = 9
\cdot 10^9$.
 The total number of events
\beq
N_P = \int^{1}_{x= 0.2} dP(x)
\label{k42}
\eeq
is also indicated in each case. The difference between the dashed and the
dotted lines shows  that
the nearby singularity in the anomaly form factor influences the decay rate
substantially at low photon energies.
The effect disappears at $x \rightarrow 1$, where $W^2 = M_K^2 (1-x)
\rightarrow 0$. To minimize the effect of resonance exchange,
the large-$x$ region should thus be
considered. The low-$x$ region, on the other hand, may be used to explore the
$W^2$ dependence of $V$ and of $A$. For a rather exhaustive discussion
of the relevance of this $W^2$ dependence
for the analysis of $K_{l2\gamma}$ decays we refer the reader to Ref.~
\cite{BARDIN}.

\subsection{Outlook}

Previous experiments have used various cuts in phase space in order (i)
to identify the individual contributions IB, {\mbox{SD}}$^\pm$, INT$^\pm$ as
far
 as
possible, and (ii) to reduce the background from $K_{l3}$ decays. This
background has in fact forced such
severe cuts that only the upper end of the
lepton spectrum remained.

The experimental possibilities to reduce the
background from $K_{l3}$ decays are
presumably more favourable with today's techniques. Furthermore, the
annual yield of $9\cdot 10^9 K^+$ decays at e.g. DAFNE is more than two orders
of magnitude higher than the samples which were available until now
 \cite{ke22,km21,ke21,km22}.
This allows for a big improvement in the determination
of the amplitudes $A$ and $V$, in particular in $K_{\mu 2 \gamma}$ decays.
 It would  be very interesting to pin
down the
combination $L_9 + L_{10}$ of the low-energy constants which occurs
in the chiral representation of the amplitude $A$, and to
investigate the $W^2$ dependence of the form factors.
\vspace{1cm}

\setcounter{equation}{0}
\setcounter{subsection}{0}
\section{The decays $K^{\pm},\;\pi^{\pm}
 \rightarrow l^{\pm} \nu_l l'^{+}l'^{-}$}
\label{KL2LL}

Here we consider decays where the photon turns into a
lepton--antilepton pair,
\begin{eqnarray}
K^+ &\to&e^+ \nu_e \mu^+\mu^- \label{S21}\\
K^+ &\to&\mu^+ \nu_\mu e^+ e^- \label{S22}\\
\pi^+ &\to&\mu^+\nu_\mu e^+ e^-\label{S25}\\
K^+ &\to&e^+ \nu_e e^+ e^- \label{S23}\\
K^+ &\to&\mu^+ \nu_\mu \mu^+ \mu^-   \label{S24}\\
\pi^+&\to&e^+ \nu_e e^+ e^-\;.\label{S26}
\end{eqnarray}

\subsection{Matrix elements}

\label{pireplace}
We start with the processes (\ref{S21}) and (\ref{S22}),
\begin{eqnarray}
K^+(p) &\to&l^+(p_l) \nu_l (p_\nu) l'^{+}(p_1) l'^{-}(p_2)\nonumber\\
(l,l')&=& \ (e,\mu)\mbox{ or } (\mu,e)\;.
\end{eqnarray}
The matrix element is
\begin{equation}
\label{S27}
T = -i G_F e V_{us}^* \overline{\epsilon}_\rho   \left\{
F_K \overline{L}^\rho - \overline{H}^{\rho\mu}l_{\mu} \right\}
\end{equation}
where
\begin{eqnarray}
\label{S28}
\overline{L}^\mu & = & m_l \overline{u}(p_\nu) (1+\gamma_5)
\left\{
\frac{2 p^\mu - q^\mu}{2 p q - q^2} -\frac{2p^\mu_l +\not\!q\gamma^\mu}
{2 p_l q + q^2}\right\}
v(p_l)
\nonumber\\[2mm]
l^\mu &=& \overline{u}(p_\nu)\gamma^\mu (1-\gamma_5) v(p_l)
\nonumber\\[2mm]
\overline{H}^{\rho\mu} & = &i V_1 \epsilon^{\rho\mu\alpha\beta}
 q_\alpha p_\beta - A_1 ( qW g^{\rho\mu} - W^\rho q^\mu )
\nonumber\\&&
-A_2 (q^2 g^{\rho\mu} - q^\rho q^\mu )
-A_4 (qW q^\rho - q^2 W^\rho) W^\mu\;,
\end{eqnarray}
with
\begin{equation}
A_4 = \frac{2 F_K}{M_K^2 - W^2} \frac{F_V^{K}   (q^2) - 1 }{q^2} + A_3
\ .
\end{equation}
The form factors $A_i(q^2,W^2),\ V_1(q^2,W^2)$ are the ones defined
in Appendix B, and
$F_V^{K}  (q^2)$ is the electromagnetic form factor
of the $K^+$. Finally the quantity $\overline{\epsilon}^{ \mu}$ stands
for
\begin{equation}
\overline{\epsilon}^{\mu} =   \frac{e}{q^2} \overline{u}(p_2)
\gamma^\mu v(p_1)\ ,
\end{equation}
and the four-momenta are
\begin{equation}
q = p_1 + p_2\  ,\ \ W = p_l + p_\nu = p - q\;,
\end{equation}
such that $q_\mu \overline{\epsilon}^{\mu} = 0$.
The first term in Eq. (\ref{S27}) contains
the part where the off-shell photon radiates off the
final-state lepton and those corrections that are reabsorbed in the
definition of $F_K$. This can always be done independently of the model
used, because of gauge invariance~\cite{Low}.

In order to obtain the matrix element for (\ref{S23}) and (\ref{S24}),
\begin{equation}
K^+ (p) \to l^+(p_l) \nu_l(p_\nu) l^+(p_1) l^-(p_2)\ ,
\end{equation}
one identifies $m_l$ and $m_l'$ in (\ref{S27}) and subtracts the
contribution obtained from interchanging $p_1 \leftrightarrow p_l$ :
\begin{eqnarray}
(p_1 , p_l) &\to& (p_l,p_1) \nonumber\\
q &\to& p_l + p_2\nonumber\\
W &\to& p- q=p_\nu + p_1\ .
\end{eqnarray}
For the decays of a pion, the same formulas with the following
replacements apply:
\begin{eqnarray}
\label{pireplace2}
M_K \to M_\pi\;,\qquad & & \qquad V_{us} \to V_{ud}\;,\nonumber\\
F_K \to F_\pi\;,\qquad & & \qquad F_V^K(q^2) \to F_V^\pi(q^2)\;.
\end{eqnarray}

\subsection{Decay distributions}

The decay width is given by
\begin{equation}
d\Gamma  =  \frac{1}{2M_K (2\pi)^8 }
\sum_{spins} |T^2| d_{LIPS}(p;p_l,p_\nu,p_1,p_2)
\end{equation}
and the total rate is the integral over this for the case $l\ne l'$.
For the case $l = l'$ the integral has to be divided by a factor of 2
for two identical particles in the final state.

We first consider the case where
$l\ne l'^{}$ and introduce the dimensionless variables
\begin{eqnarray}
x &=& \frac{2 pq}{M_K^2}\;\;,\hspace{2cm}
y  =  \frac{2p_l p}{M_K^2}\;\;,\hspace{2cm}
z  =  \frac{q^2}{M_K^2}\;\;,\nonumber\\
r_l &=& \frac{m_l^2}{M_K^2}\;\;,\hspace{2cm}
r_l'  = \frac{m_{l'}^2}{M_K^2} \; \;.
\end{eqnarray}
Then one obtains, after integrating over $p_1$ and $p_2$ at fixed $q^2$
\cite{KRISHNA},
\begin{eqnarray}
\label{S16}
d\Gamma_{K^+ \to l^+ \nu_l l'^{+}l'^{-}} &=&
\alpha^2 G_F^2 |V_{us}|^2 M_K^5 F(z,r'_l) \left\{
   -\sum_{spins} \overline{T}_\mu^* \overline{T}^\mu \right\}
    dx dy dz
\nonumber\\
F(z,r'_l)&=& \frac{1}{192 \pi^3 z} \left\{
     1 + \frac{2r'_{l}}{z}\right\}
  \sqrt{1 - \frac{4r'_{l}}{z}}
\nonumber\\
{\overline{T}}^\mu &=& M_K^{-2} \left\{
F_K \overline{L}^\mu -
\overline{H}^{\mu\nu}l_\nu\right\} .
\end{eqnarray}
The quantity $\left\{ -
\sum_{spins}\overline{T}_\mu^* \overline{T}^\mu \right\}$
can be found in Ref. \cite{BEGRep}.
The expression or a Monte Carlo program containing
this trace in FORTRAN can be obtained from the authors.
This result allows one to evaluate, e.g.
the distribution $d\Gamma/dz$ of produced $l'^+ l'^-$ pairs, rather
easily. The kinematically allowed region is
\begin{eqnarray}
4 r'_l \le&z&\le 1 + r_l - 2 \sqrt{r_l}
\nonumber\\
2 \sqrt{z} \le &x& \le 1 + z - r_l
         \\
A - B \le & y & \le A+B \;,\nonumber
\end{eqnarray}
with
\begin{eqnarray}
A&=&\frac{(2-x)(1+z+r_l -x)}{2 ( 1+z-x)}\nonumber\\
B&=&\frac{(1+z-x-r_l)\sqrt{x^2 - 4 z}}{2(1+z-x)} \; \; .
\end{eqnarray}
The case $l = l'$ is slightly more elaborate since the integration over
part of the phase space cannot be done analytically as before.
The expression for $\sum_{spins} |T|^2$,
together with the Monte Carlo program to do the phase-space integrals,
is available from the authors.

All formulas in this section are also valid with the replacements of
Eq. (\ref{pireplace2}) for the pion-decay case.

\subsection{Theory}

The form factors $A_i,\ V_1$ and $F_V^K$ have been discussed in all
kinds of models, Vector Meson Dominance, hard meson, etc.\ (see
Ref.~\cite{BARDIN} for a discussion). Here, we will
work rigorously within the framework of CHPT.

There are two classes of contributions to these decays. They  are
depicted in Fig. \ref{figure2.1}.
\begin{figure}
\vspace{7cm}
\caption{The two classes of contributions to the decays $K^+ \to
l^+ \nu_l l^{\prime +} l^{\prime -}$.}
\label{figure2.1}
\end{figure}
The first class of
diagrams is the well understood Bremsstrahlung off the
final-state lepton. This contribution is entirely \footnote{The momentum
dependence of the $W$ propagator is neglected.}
contained in the
first term in Eq. (\ref{S27}).
The second class of diagrams, which
also contains the radiation off the kaon line,
contributes to both terms in Eq. (\ref{S27}).
The one-loop diagrams contributing to this class are
shown in Fig. \ref{figure2.2}.
\begin{figure}
\vspace{7cm}
\caption{Loop diagrams for $K^+ \to W^+ \gamma^*$. The virtual photon
must be appended on all charged lines and on all vertices.}
\label{figure2.2}
\end{figure}

To leading order we have
\begin{eqnarray}
V_1&=&0 \nonumber\\
A_1\ =\ A_2&=&A_3\ = 0\ .
\end{eqnarray}
We also have $F_V^K = 1$ and $F_K$ is replaced by $F$.
The rate here is entirely given by the
inner Bremsstrahlung contribution.

The effects of the next-to-leading order CHPT corrections manifest
themselves in the replacement $F \to F_K$ and in
non-zero values for several of the form factors :
\begin{eqnarray}
V_1&=&- \frac{1}{8\pi^2 F}\nonumber\\
A_1&=&- \frac{4}{F}\left( L_9^r + L_{10}^r \right) \nonumber\\
A_2&=&- \frac{2 F_K ( F_V^K(q^2) -1 ) }{q^2}          \\
A_3&=&0\nonumber\\
F_V^K(q^2)&=&1+H_{\pi\pi}(q^2)+2 H_{KK}(q^2) \; \; . \nonumber
\label{ourresult}
\end{eqnarray}
These results obey the current algebra relation of Ref. \cite{BARDIN}.
The function $F_V^K(q^2)$ does, however, deviate somewhat from the
linear parametrization that is often used.
The function $H(t)$ is defined in Appendix \ref{notation}.
It can be easily seen from (\ref{S27}), (\ref{S28}) and (\ref{ourresult})
that the one-loop contributions to the decay matrix elements
vanish for $q^2 \to 0$ and only the contributions
from the Wess--Zumino term (to the form factor $V_1$)
and from the $p^4$ Lagrangian remain.

The fact that the form factors at
next-to-leading order could be written in terms of the kaon
electromagnetic form factor
       in a simple way is no longer true            at the $p^6$
level.
The Lagrangian at order $p^6$ contains a term of the form
\begin{equation}
\mbox{\rm tr}\left\{D_\alpha F_L^{\alpha\mu} U^{\dag} D^{\beta}
F_{R\beta\mu} U \right\}
\end{equation}
that contributes to $A_2$ and $A_3$, but not to the kaon electromagnetic
form factor,  $F_V^K (q^2)$.

Essentially
the same result holds for the decay $\pi^+ \to l^+ \nu_l e^+ e^-$
after doing the replacements (\ref{pireplace2})
in Eqs. (\ref{S27}), (\ref{S28})
and (\ref{ourresult}).
The
pion electromagnetic form factor to one loop is given
by~\cite{GLNP2,BCC}
\begin{equation}
F_V^\pi(q^2) = 1+2 H_{\pi\pi}(q^2)+ H_{KK}(q^2) \; \; .
\end{equation}
Our result for the $\pi^+ \to l^+\nu_l l^+l^-$ form factors
agrees in the limit of two flavours with the result of
Ref.~\cite{GLAP}.

\subsection{Numerical results}

Using the formulas of the previous subsections,
we have calculated the rates for a few cuts, including those
given in the literature.
The values for the masses used are those of $K^+$ and $\pi^+$.
For $L_9$ and $L_{10}$ we used the values given in Section \ref{CHPT}.

For the case of \underline{unequal} leptons, the results are given
in Table \ref{KLLL1} for the decay
$K^+ \to \mu^+ \nu_\mu e^+ e^-$. These include the cuts used in Refs.
\cite{KRISHNA} and \cite{DIAMANT},
$x \ge 40\ \mathrm{MeV} /M_K$ and $ z \ge (140\ \mathrm{MeV}/M_K)^2$,
respectively.
\begin{table}
\caption{\label{KLLL1}
         Theoretical values for the branching ratios for the decay
$K^+ \to \mu^+ \nu_\mu e^+ e^-$ for various cuts.}
\begin{center}
\begin{tabular}{|c|c|c|}
\hline
               &  tree level & form factors as given by CHPT\\
\hline
full phase space &$2.49 \cdot 10^{-5}$  &         $2.49 \cdot 10^{-5}$\\
\hline
$z\le 10^{-3}$ & $ 2.07\cdot 10^{-5} $ & $ 2.07 \cdot 10^{-5}$\\
\hline
$z\ge 10^{-3}$ &$4.12\cdot 10^{-6}$& $4.20\cdot 10^{-6}$\\
\hline
$z \ge ( 20\ MeV/M_K)^2$ & $ 3.15\cdot 10^{-6}$ & $3.23\cdot 10^{-6}$\\
\hline
$z \ge (140\ MeV/M_K)^2$ & $ 4.98\cdot 10^{-8}$ & $8.51\cdot 10^{-8}$\\
\hline
$ x\ge 40\ MeV/M_K$ & $ 1.58\cdot 10^{-5} $ & $1.58\cdot 10^{-5}$ \\
\hline
\end{tabular}
\end{center}
\end{table}
It can be seen that for this decay most of the branching ratio is
generated at very low electron--positron invariant masses. As can be seen
from the result for the cuts used in Ref. \cite{DIAMANT}, the effect
of the structure-dependent terms is most visible at high invariant
electron--positron mass. Our calculation, including the effect
of the form factors, agrees well with their data. We disagree, however,
with the numerical result obtained in Ref. \cite{KRISHNA} by about an
order of magnitude.

For the decay $K^+ \to e^+ \nu_e \mu^+\mu^-$, we obtain for the tree
level or IB contribution a branching ratio
\begin{equation}
BR_{\mathrm{IB}}(K^+ \to e^+ \nu_e \mu^+ \mu^- ) =
     3.06\cdot 10^{-12}
\end{equation}
      and,
including the form factors,
\begin{equation}
BR_{\mathrm{total}}(K^+ \to e^+ \nu_e \mu^+ \mu^- ) =
 1.12\cdot 10^{-8}.
\end{equation}
 Here the
 structure-dependent terms are the leading contribution since the inner
Brems\-strahlung  contribution is helicity-suppressed, as can be seen
from the factor $m_l$ in $\overline{L_\mu}$.

The decay $\pi^+ \to \mu^+ \nu_\mu e^+ e^-$ is entirely given by the
inner Bremsstrahlung contribution. The effects of the form factors are
very small. The IB  amplitude is not helicity-suppressed here since
$M_\mu^2 / M_\pi^2 \approx 1$. The effects of the form factors are
partly suppressed since only very small $e^+ e^-$ masses are
allowed by phase space.
The rates for full phase space and various cuts on $z$,
defined as in Eq. (\ref{defz}), but with $M_\pi$ instead of $M_K$,
are given in Table~\ref{PILLP}.
\begin{table}
\caption{\label{PILLP}
         Theoretical values for the branching ratios for the decay
$\pi^+ \to \mu^+ \nu_\mu e^+ e^-$ for various cuts. The effect of the
form factors is not visible to the accuracy quoted.}
\begin{center}
\begin{tabular}{|c|c|}
\hline
               &  tree level \\
\hline
full phase space & $3.3 \cdot 10^{-7}$ \\
\hline
$z \ge 10^{-3}$&  $6.2 \cdot 10^{-8}$\\
\hline
$z \ge 10^{-2}$&  $2.8 \cdot 10^{-9}$\\
\hline
\end{tabular}
\end{center}
\end{table}

For the decays with \underline{identical}
leptons we obtain for the muon case a
branching ratio  of
\begin{equation}
BR_{\mathrm{total}}(
K^+ \to \mu^+\nu_\mu \mu^+\mu^-) =
1.35\cdot10^{-8}
\label{predicted}
\end{equation}
for the full phase space, including the effects of the
form factors.
The inner Bremsstrahlung or the tree level branching
ratio for this decay is
\begin{equation}
BR_{\mathrm{IB}}(
K^+ \to \mu^+\nu_\mu \mu^+\mu^-) =
3.79\cdot10^{-9}.
\end{equation}
For the decay with two positrons and one electron, the integration
over the full phase space
for the tree-level results
is very sensitive to the behaviour for small
pair masses.
 We have given the tree level and the full prediction
including form-factor effects in Table \ref{KLLLP}. The cuts are
always on both invariant masses :
\begin{eqnarray}
\label{defz}
z&=&(p_1 + p_2)^2 / M_K^2 \nonumber\\
z_1 &=& (p_l + p_2)^2 / M_K^2\ .
\end{eqnarray}
\begin{table}
\caption{\label{KLLLP}
         Theoretical values for the branching ratios for the decay
$K^+ \to   e^+ \nu_e e^+ e^-$ for various cuts.}
\begin{center}
\begin{tabular}{|c|c|c|}
\hline
               &  tree level & form factors as given by CHPT\\
\hline
full phase space & $\approx 4 \cdot 10^{-9}$ & $ 1.8\cdot 10^{-7}$\\
\hline
$z,\ z_1 \ge 10^{-3}$&  $3.0\cdot 10^{-10}$&      $1.22 \cdot 10^{-7}$\\
\hline
$z,\ z_1 \ge (50\ MeV/M_K)^2 $
&  $5.2\cdot 10^{-11}$&      $8.88 \cdot 10^{-8}$\\
\hline
$z,\ z_1 \ge (140\ MeV/M_K)^2 $
&  $2.1\cdot 10^{-12}$&      $3.39 \cdot 10^{-8}$\\
\hline
\end{tabular}
\end{center}
\end{table}

For the decay $\pi^+ \to e^+\nu_e e^+e^-$, the integration over phase
space is again very sensitive to the region for small invariant masses
of the pairs.
Tree level and full results are
given for several cuts in Table~\ref{PLL};
\begin{table}
\caption{\label{PLL}
         Theoretical values for the branching ratios for the decay
$\pi^+ \to   e^+ \nu_e e^+ e^-$ for various cuts.}
\begin{center}
\begin{tabular}{|c|c|c|}
\hline
               &  tree level & form factors as given by CHPT\\
\hline
$z,\ z_1 \ge 10^{-3}$&  $2.4\cdot 10^{-9}$&      $3.0 \cdot 10^{-9}$\\
\hline
$z,\ z_1 \ge 10^{-2}$&  $4.2\cdot 10^{-10}$&      $8.7\cdot 10^{-10}$\\
\hline
Cuts similar to
\protect{\cite{scheck}}&$2.1\cdot 10^{-10}$&$5.7\cdot 10^{-10}$
\\
\hline
\end{tabular}
\end{center}
\end{table}
$z$ and $z_1$ are defined as in Eq. (\ref{defz}) with $M_K$ replaced
by $M_\pi$.
The cuts used for the last row are similar to
those of Ref.~\cite{scheck}. We require
a minimum energy of 15 MeV for the three charged particles
and an opening angle $\theta_{1,2}$
for each electron--positron pair with $\cos \theta_{1,2} \le 0.95$.

\subsection{Experimental status}

For kaon decays only those
with an electron positron pair in the final state,
decays (\ref{S22}) and (\ref{S23}),
have been observed.

Both were measured
in the same experiment       \cite{DIAMANT}.
The decay $K^+ \to \mu^+ \nu_\mu e^+ e^-$ was measured with a branching
ratio of $(1.23 \pm 0.32)\cdot 10^{-7}$ with a lower cut on the
electron--positron invariant mass of $140$ MeV.
The measurement is compatible with our calculation including the
form-factor effects for the relevant region of phase space.
This measurement
was then  extrapolated \cite{DIAMANT}
using the result of \cite{KRISHNA} to the
full phase space. Since we  disagree with that calculation,
we also disagree with the extrapolation.

In the same experiment, four events of the type
$K^+ \to e^+\nu_e e^+ e^-$
were
observed where both electron--positron pair invariant masses were
above 140 MeV. This corresponds to a branching ratio for this
region of phase space of $(2.8^{+2.8}_{-1.4})\cdot 10^{-8}$.
This result is compatible within errors  with our calculation, see
Table \ref{KLLLP}.
The matrix element of Ref. \cite{KRISHNA} was again used for the
extrapolation to full phase space \cite{DIAMANT}.
Apart from our numerical disagreement,
the calculation of Ref. \cite{KRISHNA} was for the case of
non-identical leptons and cannot be applied here.

For the decay $K^+ \to \mu^+\nu_\mu \mu^+\mu^-$ an upper limit
of $4.1 \cdot 10^{-7}$ exists \cite{ATIYA}. This upper limit is
compatible with our theoretical result, Eq. (\ref{predicted}).

The decay $K^+ \to e^+\nu_\mu \mu^+\mu^-$ has not been looked for so far
but should be within the capabilities of a $\Phi$ factory,
given the
branching ratio predicted in the previous subsection.
This decay proceeds
almost entirely through the structure-dependent terms and, as such, is a
good test of our calculation.

The decay $\pi^+ \to \mu^+\nu_\mu e^+e^-$ has not yet been looked for.
It is almost entirely given by
the inner Bremsstrahlung contribution so
it is not very interesting as a test of the form factors. The decay
$\pi^+ \to e^+\nu_e e^+e^-$
has been measured in \cite{egli}.
We do not make a direct comparison with our result,
because the acceptance
of the apparatus is very difficult to implement,
see Ref. \cite{Thesis}.
We do agree with the form factors as determined there.

\subsection{Outlook}

The decays discussed in this section, $K^+ \to l^+\nu_l
l'^+l'^-$, are
complementary to the decays $K^+ \to l^+\nu_l\gamma$. As was the case
for the analogous decay $\pi^+ \to e^+\nu_e e^+ e^-$ \cite{egli},
 it may be
possible to explore phase space
more easily with this process than with
$K^+ \to l^+ \nu_l \gamma$ to resolve ambiguities in the form factors.

As can be seen from our predictions, Tables \ref{KLLL1} and
\ref{KLLLP}, all the decays considered in this section should be
observable at a $\Phi$ factory like DAFNE or in fixed-target experiments
at future kaon factories.
Large improvements in statistics are
possible since less severe cuts than those used in the past
experiments should be possible. In the decays with a $\mu^+ \mu^-$
pair and the decay $K^+ \to e^+ \nu_e e^+ e^-$,
the effects of the form factors are already large in the total
rates. In the decay $K^+ \to \mu^+ \nu_\mu e^+ e^-$
most of the total rate is for a small invariant
mass of the pair and is given by the inner Bremsstrahlung
contribution.  There are, however, regions of phase space where the
form-factor effects are large and where branching ratios are still
large enough that sufficient statistics can be accumulated.
\vspace{1cm}

\setcounter{equation}{0}
\section{Radiative $K_{l3}$ decays} \label{Kl3g}
The decay channels considered in this section are
\beqa
K^+(p) & \ra & \pi^0(p') l^+(p_l) \nu_l(p_{\nu}) \gamma(q) \qquad
[K^+_{l3\gamma}] \no \\*
K^0(p) & \ra & \pi^-(p') l^+(p_l) \nu_l(p_{\nu}) \gamma(q) \qquad
[K^0_{l3\gamma}] \no
\eeqa
and the charge conjugate modes. We only consider real photons ($q^2 = 0$).

\subsection{Kinematics and invariant amplitudes}
The matrix element for $K^+_{l3\gamma}$ has the general structure
\beqa
T & = & \left.\dfrac{G_F}{\sqrt{2}} e V^*_{us} \ve^{\mu}(q)^*
\right\{(V^+_{\mu\nu} -
A^+_{\mu\nu}) \ol{u}(p_{\nu}) \gamma^{\nu} (1 - \gamma_5)
v(p_l) \label{eq:T}  \\*
  &   & + \left. \dfrac{F^+_{\nu}}{2 p_lq} \ol{u}(p_{\nu}) \gamma^{\nu}
(1 - \gamma_5) (m_l - \fsl p_l - \fsl q) \gamma_{\mu} v(p_l)\right\}
\equiv \ve^{\mu *} A^+_{\mu}. \no
\eeqa
The diagram of Fig. \ref{fig51}a, corresponding to the first part of
Eq. (\ref{eq:T}), includes Bremsstrahlung off the $K^+$.
The lepton Bremsstrahlung diagram of Fig. \ref{fig51}b is represented
by the second part of Eq. (\ref{eq:T}).
The hadronic tensors $V^+_{\mu\nu}, A^+_{\mu\nu}$ are defined as
\beq
I^+_{\mu\nu} = i \int d^4x e^{i q x} \langle \pi^0(p') \mid
T\{V^{em}_\mu(x) I^{4-i 5}_\nu(0)\} \mid K^+(p) \rangle\; ,
\qquad I = V,A \;;
\eeq
$F^+_\nu$ is the $K^+_{l3}$ matrix element
\beq
F^+_\nu = \langle \pi^0(p') \mid V^{4-i 5}_\nu(0) \mid K^+(p) \rangle\;.
\eeq
The tensors $V^+_{\mu\nu}$ and $A^+_{\mu\nu}$ satisfy the Ward identities
(cf. Appendix C)
\beqa
q^\mu V^+_{\mu\nu} & = & F^+_\nu \label{eq:Ward} \\*
q^\mu A^+_{\mu\nu} & = & 0\;,  \no
\eeqa
leading in turn to
\beq
q^\mu A^+_\mu = 0~, \label{eq:WI}
\eeq
as is required by gauge invariance.

For $K^0_{l3\gamma}$, one obtains the corresponding amplitudes and
hadronic tensors by making the replacements
\beqa
K^+ & \ra & K^0\;,\qquad \pi^0 \ra \pi^- \;,\no \\*
V^+_{\mu\nu} & \ra & V^0_{\mu\nu}\;,
\qquad A^+_{\mu\nu} \ra A^0_{\mu\nu}\;,\\*
F^+_\nu & \ra & F^0_\nu\;,\qquad A^+_\mu \ra A^0_\mu\;. \no
\eeqa

To make the infrared behaviour transparent,
it is convenient to separate the tensors $V^+_{\mu\nu}, V^0_{\mu\nu}$
into two parts:
\beqa
V^+_{\mu\nu} & = & \hat{V}^+_{\mu\nu} + \dfrac{p_\mu}{pq} F^+_\nu
\label{eq:Low}\;, \\*
V^0_{\mu\nu} & = & \hat{V}^0_{\mu\nu} + \dfrac{p'_\mu}{p'q} F^0_\nu\;.
 \no
\eeqa
Due to Low's theorem \cite{Low}, the amplitudes $\hat{V}^{+,0}_{\mu\nu}$
are finite for $q \ra 0$. The axial amplitudes
$A^{+,0}_{\mu\nu}$ are automatically infrared-finite.
The Ward identity (\ref{eq:Ward}) implies that the vector amplitudes
$\hat{V}^{+,0}_{\mu\nu}$ are transverse:
\beq
q^\mu \hat{V}^{+,0}_{\mu\nu} = 0\;. \eeq

For on-shell photons, Lorentz and parity invariance together with gauge
invariance allow the general
decomposition (dropping the superscripts +, 0 and terms that vanish
upon contraction with the photon polarization vector)
\beqa
\hat{V}_{\mu\nu} & = & V_1 \left(g_{\mu\nu} - \dfrac{W_\mu q_\nu}
{qW}\right) + V_2 \left(p'_\mu q_\nu - \dfrac{p'q}{qW} W_\mu q_\nu
\right) \\* \no
& & + V_3 \left( p'_\mu W_\nu - \dfrac{p'q}{qW} W_\mu W_\nu \right)
+ V_4\left( p'_\mu p'_\nu - \dfrac{p'q}{qW} W_\mu p'_\nu \right)
\label{eq:tensor} \\
A_{\mu\nu} & = & i \ve_{\mu\nu\rho\sigma} (A_1 p'^\rho q^\sigma +
A_2 q^\rho W^\sigma) + i \ve_{\mu\lambda\rho\sigma} p'^\lambda
q^\rho W^\sigma (A_3 W_\nu + A_4 p'_\nu) \no \\
F_\nu & = & C_1 p'_\nu + C_2 (p - p')_\nu  \no \\
W & = &  p_l + p_\nu\;. \no
\eeqa
For the decomposition of the axial tensor amplitude, use was made
of Schouten's identity
\beq
\sum_{(\lambda\mu\nu\rho\sigma)} v_\lambda \ve_{\mu\nu\rho\sigma} = 0\;,
\eeq
where the sum extends over all cyclic permutations of the five indices
and $v$ is an arbitrary four-vector.
With the decomposition (\ref{eq:Low}) we can write the matrix element
for $K^+_{l3\gamma}$ in (\ref{eq:T}) in a form analogous to Eqs.
(\ref{k3},\ref{k4}) for $K_{l2\gamma}$:
\beqa
T & = & \left.\dfrac{G_F}{\sqrt{2}} e V^*_{us} \ve^{\mu}(q)^*
\right\{(\hat{V}^+_{\mu\nu} -
A^+_{\mu\nu}) \ol{u}(p_{\nu}) \gamma^{\nu} (1 - \gamma_5) v(p_l)
\label{eq:Tnew}  \\*
  &   & + \left. F^+_{\nu} \ol{u}(p_{\nu}) \gamma^{\nu}
(1 - \gamma_5) \left[ \dfrac{p_\mu}{pq} -
\dfrac{2 p_{l\mu} + \fsl q \gamma_\mu}
{2 p_lq} \right] v(p_l)\right\}~. \no
\eeqa

The four invariant vector amplitudes $V_1,\ldots,V_4$ and
the four axial amplitudes $A_1,\ldots,A_4$ are functions of three scalar
variables. A convenient choice for these variables is
\beq
E_\gamma = pq/M_K,\qquad \; E_\pi = pp'/M_K ,\; \qquad
W = \sqrt{W^2}\;,
\label{eq:kin1} \eeq
where $W$ is the invariant mass of the lepton pair. The amplitudes
$C_1, C_2$ can be expressed in terms of the $K_{l3}$ form factors
and depend only on the variable
$ (p - p')^2 = M^2_K + M^2_\pi - 2 M_K E_\pi$.
\begin{figure}[t]
\vspace{7cm}
\caption{Diagrammatic representation of the $K^+_{l3\gamma}$
amplitude.} \label{fig51}
\end{figure}
For the full kinematics of $K_{l3\gamma}$ two more variables are
needed, e.g.
\beq E_l = pp_l/M_K ,\; x = p_lq/M^2_K \;. \label{eq:kin2} \eeq
The variable $x$ is related to the angle $\theta_{l\gamma}$ between the
photon and the charged lepton in the $K$ rest frame:
\beq
x M^2_K = E_\gamma (E_l - \sqrt{E^2_l - m^2_l} \cos{\theta_{l\gamma}})
\;.
\eeq
T invariance implies that the vector amplitudes $V_1,\ldots,
V_4$, the
axial amplitudes $A_1,\ldots,A_4$ and the $K_{l3}$ form factors
$C_1, C_2$ are (separately) relatively real in the physical region.
We choose the standard
phase convention in which all amplitudes are real.

For $\theta_{l\gamma} \ra 0$ (collinear lepton and photon),
there is a lepton-mass singularity in (\ref{eq:T}),
which is numerically
relevant for $l = e$ \cite{Doncel}.
The region of small $E_\gamma, \theta_{l\gamma}$ is dominated by the
$K_{l3}$ matrix elements. The new theoretical information of
$K_{l3\gamma}$ decays resides in the infrared-finite tensor amplitudes
$\hat{V}_{\mu\nu}$ and $A_{\mu\nu}$.
The relative importance of these
contributions can be enhanced by cutting away the region of low
$E_\gamma, \theta_{l\gamma}$.

\subsection{CHPT to $O(p^4)$}
Prior to CHPT, the most detailed calculations of $K_{l3\gamma}$
amplitudes were performed by Fearing, Fischbach and Smith
\cite{FFS} using current-algebra techniques.

To lowest order in the chiral expansion [$O(p^2)$], the invariant
amplitudes for the two charge modes are given
by \cite{Hol}
\nopagebreak
\begin{flushleft} $\ul{K^+_{l3\gamma}:}$ \end{flushleft}
\beqa
V^+_1 & = & \dfrac{1}{\sqrt{2}} \no \\*
V^+_2 & = & - \dfrac{1}{\sqrt{2}pq} \\
C^+_1 & = & 2 C^+_2 \enskip = \enskip \sqrt{2} \no \eeqa
\nopagebreak \begin{flushleft} $\ul{K^0_{l3\gamma}:}$ \end{flushleft}
\beqa
V^0_1 & = & - 1  \no \\*
V^0_2 & = & \dfrac{1}{p'q} \\*
C^0_1 & = & 2 C^0_2 \enskip = \enskip 2 \no
\eeqa
All other invariant amplitudes vanish to $O(p^2)$.

At next-to-leading order in CHPT,
there are in general three types of contributions~\cite{GLNP1}:
anomaly, local contributions due to ${\cal L}_4$, and loop amplitudes.
\begin{figure}
\vspace{7.5cm}
\caption{Loop diagrams (without tadpoles) for $K_{l3}$. For
$K_{l3\gamma}$, the photon must be appended on all charged lines and
on all vertices.} \label{fig52}
\end{figure}
The chiral anomaly encoded in the WZW functional yields
the axial amplitudes  \beqa
A^+_{\mu\nu} & = & \dfrac{i \sqrt{2}}{16 \pi^2 F^2} \left\{\ve_
{\mu\nu\rho\sigma}q^\rho(4 p'+W)^\sigma + \dfrac{4}{W^2-M^2_K}
\ve_{\mu\lambda\rho\sigma}W_\nu p'^\lambda q^\rho W^\sigma \right\}
\label{eq:anom} \\*
A^0_{\mu\nu} & = & - \dfrac{i}{8 \pi^2 F^2}\ve_{\mu\nu\rho\sigma}
q^\rho W^\sigma.\no \eeqa
The local parts of these amplitudes can be read off the
Lagrangian (\ref{eq:semi}). The non-local kaon pole term in
$A^+_{\mu\nu}$ is due to the anomalous $K^+ K^- \pi^0 \gamma$ vertex
contained in (\ref{eq:ang}) with the outgoing $K^+$ turning into
a $W^+$. Its contribution to the matrix element is helicity-suppressed
\cite{FFS,Hol} and can be neglected for $l=e$. However, for
$K^+_{\mu3\gamma}$ it is a normal-size $O(p^4)$ contribution to the
axial amplitude.

The loop diagrams for $K_{l3\gamma}$ are shown in
Fig. \ref{fig52}. We first write the $K^+_{l3}$ matrix element
in terms of three functions $f^+_1, f^+_2,f^+_3$ which will also appear
in the invariant amplitudes $V^+_i$. Including the contributions
from the low-energy constants $L_5,L_9$ in ${\cal L}_4$, the
$K_{l3}$ matrix element $F^+_\nu$ is given by \beqa
C^+_1 & = & f^+_1(t) \no \\*
C^+_2 & = & \dfrac{1}{2}(M^2_K - M^2_\pi - t)f^+_2(t)
+ f^+_3(t) \no \\
f^+_1(t) & = & \sqrt{2} + \dfrac{4 \ol{L_9}}{\sqrt{2} F^2} t +
2 \sum_{I=1}^{3} (c^I_2 - c^I_1) B^I_2(t) \no \\
f^+_2(t) & = & - \dfrac{4 \ol{L_9}}{\sqrt{2} F^2} + \dfrac{1}{t}
\sum_{I=1}^{3} \left\{ (c^I_1 - c^I_2)\left [2 B^I_2(t)
- \dfrac{(t+\Delta_I)
\Delta_I J_I(t)}{2 t}\right ] - c^I_2 \Delta_I J_I(t)
\right\} \no \\
f^+_3(t) & = & \dfrac{F_K}{\sqrt{2} F_\pi} + \dfrac{1}{2 t}
\sum_{I=1}^{3} \left\{(c^I_1+c^I_2)(t+\Delta_I)
- 2 c^I_3\right\}\Delta_I J_I(t) \label{eq:fi} \\
\ol{L_9} & = & L^r_9(\mu) - \dfrac{1}{256 \pi^2} \ln{\dfrac
{M_\pi M^2_K M_\eta}{\mu^4}}  \no \\
\Delta_I & = & M^2_I-m^2_I\enskip ,\enskip t \enskip =\enskip (p-p')^2.
\no \eeqa
$\ol{L_9}$ is a scale-independent coupling constant and we have traded
the tadpole contribution together with $L_5$ for $F_K/F_\pi$ in
$f^+_3(t)$ [cf. Eq. (\ref{eq:FK})].
The sum over $I$ corresponds to the three one-loop diagrams
of Fig. \ref{fig52}, with the
coefficients $c^I_1,c^I_2,c^I_3$ displayed in
Table~\ref{tab:loop+}.
\begin{table} \centering
\caption{Coefficients for the $K^+_{l3\gamma}$ loop amplitudes
corresponding to the diagrams $I=1,2,3$ in Fig. \protect\ref{fig52}.
All coefficients $c^I_i$ must be divided by
$6 \protect\sqrt{2} F^2$.}
\label{tab:loop+} \vspace{.5cm}
$\ba{|l|l|l|r|r|c|}
\hline
I & M_I & m_I & c^I_1
& c^I_2  & c^I_3 \\ \hline
1 & M_K & M_\pi & 1 & -2 & - M^2_K - 2 M^2_\pi \\
2 & M_K & M_\eta & 3 & -6 & - M^2_K - 2 M^2_\pi \\
3 & M_\pi & M_K & 0 & -6 & -6 M^2_\pi \\
\hline
\ea$ \end{table}
We use the Gell-Mann--Okubo mass formula
throughout to express $M^2_{\eta}$ in terms of $M^2_K, M^2_\pi$.
The functions $J_I(t)$ and $B^I_2(t)$ can be found in
Appendix \ref{notation}.

The standard $K_{l3}$ form factors $f_+(t), f_-(t)$ are \cite{GLNP2}
\beqa
f_+(t) & = & \dfrac{1}{\sqrt{2}}f^+_1(t) \\*
f_-(t) & = & \dfrac{1}{\sqrt{2}}\left[(M^2_K-M^2_\pi-t)f^+_2(t)
+2 f^+_3(t)-f^+_1(t)\right]. \no \eeqa

It remains to calculate the infrared-finite tensor amplitude
$\hat{V}^+_{\mu\nu}$. The invariant amplitudes $V^+_i$ can be
expressed in terms of the previously defined functions $f^+_i$ and of
additional amplitudes $I_1,I_2,I_3$. Diagrammatically, the latter
amplitudes arise from those diagrams in Fig.~\ref{fig52} where the
photon is not appended on the incoming $K^+$ (non-Bremsstrahlung
diagrams). The final expressions are
\beqa
V^+_1 & = & I_1 + p'W_q f^+_2(W^2_q) + f^+_3(W^2_q)
\no \\*
V^+_2 & = & I_2 -\dfrac{1}{pq}\left[p'W_q f^+_2(W^2_q)
+ f^+_3(W^2_q)\right]
\no \\*
V^+_3 & = & I_3 +\dfrac{1}{pq}\left[p'W f^+_2(W^2)+f^+_3(W^2)
-p'W_q f^+_2(W^2_q) - f^+_3(W^2_q)\right]  \label{eq:Vhat} \\*
V^+_4 & = & \dfrac{f^+_1(W^2)-f^+_1(W^2_q)}{pq}  \no \\*
W_q & = & W + q \enskip = \enskip p - p'~.
\no \eeqa

The amplitudes $I_1, I_2, I_3$ in Eq.~(\ref{eq:Vhat}) are given by
\beqa
I_1 & = & \dfrac{4 qW}{\sqrt{2}F^2}(\ol{L_9}+\ol{L_{10}}) + \dfrac
{8 p'q}{\sqrt{2}F^2}\ol{L_9}  \no \\
&  & +\sum_{I=1}^{3}\left\{\left [(W^2_q+\Delta_I)(c^I_1+c^I_2)-
2(c^I_2 p'W_q+c^I_3)\right ]
\left[\dfrac{(W^2_q-\Delta_I)\hat{J}_I}
{2 W^2_q} -2 G_I) \right] \right.\no \\
& & +\dfrac{(c^I_2-c^I_1)}{2}
\left [\dfrac{p'W_q}{W^2_q}(\dfrac{(W^4_q-\Delta^2_I)
\hat{J}_I}{W^2_q}+4\hat{B}^I_2)+p'(W-q)L^I_m\right ] \no \\
& & +\dfrac{2(c^I_2-c^I_1)}{qW}\left.\left [p'q(F_I-
(W^2_q+\Delta_I)G_I)+p'W(\hat{B}^I_2-B^I_2)\right ] \right\}
\label{eq:loop} \no \\
I_2 & = & -\dfrac{8\ol{L_9}}{\sqrt{2}F^2}+\dfrac{2}{qW}\sum_{I=1}
^{3}(c^I_2-c^I_1)\left [F_I-(W^2+\Delta_I)G_I\right ]  \\
I_3 & = & -\dfrac{4\ol{L_9}}{\sqrt{2}F^2}+\sum_{I=1}^{3}
\left\{2(c^I_2-c^I_1)
\left [G_I+\dfrac{L^I_m}{4}+\dfrac{\hat{B}^I_2-B^I_2}{qW}
\right ] -c^I_1 \dfrac{\Delta_I J_I}{W^2} \right\} \no \\
\ol{L_{10}} & = & L^r_{10}(\mu)+\dfrac{1}{256\pi^2}\ln{\dfrac{M_\pi
M^2_K M_\eta}{\mu^4}}  \no \\
L^I_m & = & \dfrac{\Sigma_I}{32\pi^2\Delta_I}
\ln{\dfrac{m^2_I}{M^2_I}} \no \\
F_I & = & \hat{B}^I_2-\dfrac{W^2}{4}L^I_m+\dfrac{1}{qW}
\left(W^2 B^I_2-W^2_q \hat{B}^I_2\right) \no \\
G_I & = & \dfrac{M^2_I}{2}C(W^2_q,W^2,M^2_I,m^2_I)+\dfrac{1}{8 qW}
\left[(W^2_q+\Delta_I)\hat{J}_I-(W^2+\Delta_I)J_I\right]
+\dfrac{1}{64\pi^2} \no \\
J_I & \equiv & J_I(W^2), \; \hat{J}_I \equiv J_I(W^2_q) \no \\
B^I_2 & \equiv & B^I_2(W^2), \; \hat{B}^I_2 \equiv B^I_2(W^2_q)\;.
 \no \eeqa

The function $C(W^2_q,W^2,M^2_I,m^2_I)$ is given in Appendix \ref{notation}.
All the invariant amplitudes $V^+_1,\ldots,V^+_4$ are
real in the physical region. Of course, the same is true for the
$K_{l3}$ form factors $C^+_1,C^+_2$.

The $K^0_{l3\gamma}$ amplitude has a very similar structure. Both the
$K^0_{l3}$ matrix element $F^0_\nu$ and the infrared-finite
vector amplitude $\hat{V}^0_{\mu\nu}$ can be obtained from the
corresponding quantities $F^+_\nu$  and
$\hat{V}^+_{\mu\nu}$ by the following steps:
\bit
\item interchange $p'$ and $-p \:$ ;
\item replace $\dfrac{F_K}{F_\pi}$ by $\dfrac{F_\pi}{F_K}$ in $f^+_3$;
\item insert the appropriate coefficients $c^I_i$ for
$K^0_{l3\gamma}$ listed in Table~\ref{tab:loop0};
\item multiply $F^+_\nu$ and $\hat{V}^+_{\mu\nu}$ by a
factor $-\sqrt{2}$.
\eit
\begin{table} \centering
\caption{Coefficients for the $K^0_{l3\gamma}$ loop amplitudes
corresponding to the diagrams $I=1,2,3$ in Fig. \protect\ref{fig52}. All
coefficients $c^I_i$ must be divided by $6 \protect\sqrt{2}F^2$.}
\label{tab:loop0} \vspace{.5cm}
$\ba{|l|l|l|r|r|c|}
\hline
I & M_I & m_I  & c^I_1
& c^I_2 & c^I_3 \\ \hline
1 & M_K & M_\pi & 0 & -3 & -3 M^2_K \\
2 & M_K & M_\eta & 6 & -3 &  M^2_K+2 M^2_\pi \\
3 & M_\pi & M_K & 4 & -2 & -2 M^2_K + 2 M^2_\pi \\
\hline \ea $
\end{table}

\subsection{Numerical results}

The total decay rate is given by \beqa
\Gamma(K \ra \pi l \nu_l \gamma) & = & \dfrac{1}{2 M_K (2 \pi)^8}
\int d_{LIPS}(p;p',p_l,p_\nu,q)
\sum_{spins}\mid T \mid^2 \label{eq:PS}
\eeqa
in terms of the amplitude $T$ in (\ref{eq:T}). The square of the matrix
element, summed over photon and lepton polarizations, is a bilinear form
in the invariant amplitudes $V_1,\ldots,V_4$,
$A_1,\ldots,A_4$,$C_1,C_2$. The explicit expression can be found in
Ref. \cite{BEGRep}, which also contains a summary of the presently
available data sample.

Detailed experimental studies of $K_{l3\gamma}$ decays will become
feasible with the high kaon fluxes of various factories, which are either
under construction or still in the planning stage. The following number
of events corresponds to the kaon
fluxes expected in the proposed detector
KLOE \cite{Franz} for the $\Phi$ factory DAFNE in Frascati. These fluxes
are based on a luminosity of $5\times 10^{32}$ cm$^{-2}$ s$^{-1}$
equivalent to an annual rate of $~9\times 10^9 ~(1.1\times 10^9)$
tagged $K^\pm~(K_L)$ assuming a year of $10^7$ s.

For the calculation of decay rates we have used the following cuts:
\beqa
E_\gamma & \geq & 30\ {\mathrm{MeV}} \\*
\theta_{l\gamma} & \geq & 20^\circ. \no \eeqa
The physical values of $M_\pi$ and $M_K$ are used in the amplitudes;
$M_\eta$ is calculated from the Gell-Mann--Okubo mass formula. The
values of the other parameters can be found in Sect.~\ref{CHPT}.
For $K^0_{l3\gamma}$, the rates are to be understood as
$\Gamma(K_L \ra \pi^{\pm} l^{\mp} \nu \gamma)$.

The results for $K^+_{l3\gamma}$ and $K^0_{l3\gamma}$ are displayed in
Tables \ref{tab:+} and \ref{tab:0}, respectively. For comparison, the
tree-level branching ratios and the rates for
the amplitudes without the loop contributions are also shown. The
separation between loop and counter\-term contributions is of course
scale-dependent. This scale dependence is absorbed in the scale-invariant
constants $\ol{L_9}, \ol{L_{10}}$ defined in Eqs.~(\ref{eq:fi}),
(\ref{eq:loop}). In other words, the entries in Tables \ref{tab:+},
\ref{tab:0} for the amplitudes without loops correspond to setting
all coefficients $c^I_i$ in Tables \ref{tab:loop+},
\ref{tab:loop0} equal to zero.

\begin{table} \centering
\caption{Branching ratios and expected number of events at DAFNE
for $K^+_{l3\gamma}$.} \label{tab:+} \vspace{.5cm}
$\ba{|l|c|c|}
\hline
\multicolumn{1}{|c|}{K^+_{e3\gamma}} &
$BR$  & \# $events/yr$ \\ \hline
$full $ O(p^4) $ amplitude$ & 3.0\times 10^{-4} & 2.7\times 10^6 \\
$tree level$ & 2.8\times 10^{-4} & 2.5\times 10^6 \\
O(p^4) $ without loops$ & 3.2\times 10^{-4} & 2.9\times 10^6 \\
\hline \ea$ \\*[.5cm] $\ba{|l|c|c|} \hline
\multicolumn{1}{|c|}{K^+_{\mu 3\gamma}}  &
$BR$ & \# $events/yr$ \\ \hline
$full $ O(p^4) $ amplitude$ & 2.0\times 10^{-5} & 1.8\times 10^5 \\
$tree level$ & 1.9\times 10^{-5} & 1.7\times 10^5 \\
O(p^4) $ without loops$ & 2.1\times 10^{-5} & 1.9\times 10^5 \\
\hline \ea$ \end{table}
\nopagebreak[2]
\begin{table} \centering
\caption{Branching ratios and expected number of events at DAFNE
for $K^0_{l3\gamma}$.} \label{tab:0} \vspace{.5cm}
$\ba{|l|c|c|}
\hline
\multicolumn{1}{|c|}{K^0_{e3\gamma}}  &
$BR$  & \# $events/yr$ \\ \hline
$full $ O(p^4) $ amplitude$ & 3.8\times10^{-3} & 4.2\times 10^6 \\
$tree level$ & 3.6\times 10^{-3} & 4.0\times 10^6 \\
O(p^4) $ without loops$ & 4.0\times 10^{-3} & 4.4\times 10^6 \\
\hline \ea$ \\*[.5cm] $\ba{|l|c|c|} \hline
\multicolumn{1}{|c|}{K^0_{\mu 3\gamma}}  &
$BR$ & \#$events/yr$ \\ \hline
$full $ O(p^4) $ amplitude$ & 5.6\times 10^{-4} & 6.1\times 10^5 \\
$tree level$ & 5.2\times 10^{-4} & 5.7\times 10^5 \\
O(p^4) $ without loops$ & 5.9\times 10^{-4} & 6.5\times 10^5 \\
\hline \ea$ \end{table}

These numerical results show very clearly that the non-trivial
CHPT effects of $O(p^4)$ can be detected at dedicated machines such as
DAFNE, in all four channels,
without any problem of statistics. Of course, the rates are bigger for
the electronic modes. On the other hand, the relative size of the
structure-dependent terms is somewhat bigger in the muonic channels
(around 8\% for the chosen cuts). We observe that there is negative
interference between the loop and counterterm amplitudes. With the
chosen distinction between loop and counterterm contributions discussed
above, the loops amount to approximately 50\% of the
rate due to the local amplitudes of $O(p^4)$ (including the anomalous
ones).

The sensitivity
 to the counterterm coupling constants $L_9, L_{10}$ and to the
chiral anomaly can be expressed as the difference in the number of events
between the tree level and the $O(p^4)$ amplitudes (without loops).
In the optimal case of $K^0_{e3\gamma}$, this amounts to more than
$4 \times 10^5$ events/yr at DAFNE. Most of this difference is
due to $L_9$.

The chiral anomaly is more important for $K^+_{l3\gamma}$, but even there
it influences the total rates rather little. A dedicated
study of differential rates is necessary to locate the chiral anomaly in
$K_{l3\gamma}$ amplitudes, if this is at all possible.

However, taking
into account that both $L_9$ and $L_{10}$ are already known to rather
good accuracy (see Sect. \ref{CHPT}), $K_{l3\gamma}$
decays will certainly allow for precise tests of CHPT to $O(p^4)$.
\vspace{1cm}

\section{Conclusions}
\ben
\item
In this paper we have calculated the matrix elements
for the radiative semileptonic kaon decays $K^+ \to l^+ \nu_l \gamma$,
$l^+ \nu_l l'^+ l'^-$ and $K \to \pi l \nu_l \gamma (l,l'=e,\mu)$ to
next-to-leading order in CHPT.
All measured \cite{PDG}  semileptonic $K$ decay amplitudes are now known
\cite{GLNP2,BEGRep,Bijnens,Rigg} to order $p^4$ in CHPT, except for one
form factor in $K_{\mu4}$ decays \cite{coga}.

\item
The $K_{l2\gamma}$ decay does not receive any contribution
from the one-loop corrections, except for those that can be absorbed
into $F_K$. This decay is then completely predicted,
including effects of $SU(3)$ breaking,
from the knowledge
of the $\gamma$ parameter as measured in $\pi^+ \to e^+\nu_e \gamma$.
This prediction was successfully confronted with the present
experimental data, and the relevance of more detailed measurements at
future kaon facilities was discussed.

\item
For the decays with a lepton--antilepton pair in the final state
($K^+(\pi^+) \to l^+ \nu_l l'^+ l'^-$), the additional effects of
one-loop contributions can be described by the electromagnetic form
factor. It was shown that this relation breaks down at order $p^6$.
We noted that decays with a $\mu^+\mu^-$ pair are especially
useful probes of the form factors. All these decays should be measurable
in the near future with sufficient statistics to measure the form
factors.

\item
The last considered decay, $K_{l3\gamma}$, is a useful test of
the non-trivial aspects of CHPT since the
one-loop effects are of the same magnitude as the effects
of the tree-level $O(p^4)$ action. All form factors vary
rather significantly over the available kinematical region, so that an
analysis of experiments with a constant form factor will not be able
to uncover these effects. We therefore urge future experiments to
perform the analysis of their data directly with the formulas given here.

\item
Overall, the agreement with the presently available data is quite
satisfactory. On the other hand, future kaon facilities like DAFNE
will have the opportunity to test the predictions of CHPT at
next--to--leading order in much more detail than is possible with
present data.
\een
\nopagebreak \noindent  {\bf Acknowledgements}

\noindent
We thank the members of the DAFNE Working Groups, in particular
L.~Maiani and N.~Paver, for many discussions and valuable
suggestions at various stages of this work. We are grateful to
J.~Beringer for checking traces in $K_{l2\gamma}$, and to
E.A.~Ivanov for making available to us
unpublished work by D.Yu.~Bardin and collaborators on radiative pion and
kaon decays.
We have enjoyed very useful discussions with S.~Egli and C.~Niebuhr
concerning various aspects of $\pi^+ \to e^+ \nu_e e^+ e^-$.

\appendix

\newcounter{zahler}
\renewcommand{\thesection}{\Alph{zahler}}
\renewcommand{\theequation}{\Alph{zahler}.\arabic{equation}}
\setcounter{zahler}{1}

\newpage
\setcounter{equation}{0}

\noindent
{\large \bf APPENDICES}
\section{Notation and loop integrals}
\label{notation}

The notation for phase space is the one without the factors of $2\pi$.
For the decay rate of a
particle with four-momentum $p$ into $n$ particles
with momenta $p_1,\ldots,p_n$, this is
\begin{equation}
d_{LIPS}(p;p_1,\ldots,p_n) = \delta^4\left(p-\sum_{i=1}^n p_i \right)
\prod_{i=1}^n \frac{d^3 p_i}{2 p_i^0} ~.
\end{equation}
We use a covariant normalization of one-particle states,
\beq
\left\langle{\vec{p}} \; '|\vec{p}\right\rangle
= (2\pi)^3 2p^0 \delta^3({\vec{p}} \; '-\vec{p})
 \;
\; , \eeq
together with the spinor normalization
\beq
\bar{u}(p,r)u(p,s) = 2m\delta_{rs} \; \; .
\eeq
The kinematical function $\lambda(x,y,z)$ is defined as
\begin{equation}
\lambda(x,y,z) = x^2 + y^2 + z^2 - 2 (xy + yz + zx)~.
\end{equation}
We take the standard model in the current $\times$ current form, i.e.
we neglect the momentum dependence of the $W$ propagator. The currents
used in the text are :
\begin{eqnarray}
V_\mu^{4-i5} &=& \bar{q} \gamma_\mu \frac{1}{2}(\lambda_4 - i\lambda_5)q
  ~=~ \overline{s}\gamma_\mu u
\nonumber\\
A_\mu^{4-i5} &=& \bar{q} \gamma_\mu \gamma_5 \frac{1}{2}
(\lambda_4 - i\lambda_5) q
  ~=~ \overline{s}\gamma_\mu \gamma_5 u
\nonumber\\
V^{em}_\mu &=& \bar{q}\gamma_\mu Q q
\nonumber\\
Q&=&\mbox{diag}(2/3, -1/3 ,-1/3)~.
\end{eqnarray}
The numerical values used in the programs are the physical masses
for the particles as given by the Particle Data Group \cite{PDG}.
In addition we have used the values for the decay constants derived
from the most recent measured charged pion and kaon
semileptonic decay rates \cite{PDG,lroos}:
\begin{eqnarray}
F_\pi &=& 93.2 ~\mathrm{MeV}\nonumber\\
F_K &=& 113.6~\mathrm{MeV}\;.
\end{eqnarray}
We do not need values for the quark masses. For the processes considered
in this paper we can always use the lowest-order relations to rewrite
them in   terms of the
pseudoscalar meson masses (see Section \ref{CHPT}). For the
KM matrix element $|V_{us}|$
we used the central value, 0.220, of Ref. \cite{PDG}.
The numerical values for the $L_i^r(M_\rho)$ are those given in
Section \ref{CHPT}.

Whenever we quote a branching ratio for a semileptonic
$K^0$ decay, it stands for the branching ratio of the corresponding
$K_L$ decay, e.g.
\begin{equation}
BR(K^0 \to \pi^- l^+ \nu_l ) \equiv BR(K_L \to \pi^{\pm} l^{\mp} \nu )~.
\end{equation}
We use the Condon--Shortley phase conventions throughout.

Next we define the functions appearing in the loop integrals
used in the text.
First we define those needed for loops with two propagators,
mainly in the form given in Ref.~\cite{GLNP1}.
We consider a loop with two masses, $M$ and $m$.
 All needed functions can be given
in terms of the subtracted scalar integral $\bar{J}(t) = J(t) - J(0)$,
\begin{equation}
J(t) = ~  -i
\int \frac{d^4p}{(2\pi)^d} \frac{1}{((p+k)^2 - M^2)(p^2 - m^2)}
\end{equation}
with $t = k^2$.
The functions used in the text are then :
\begin{eqnarray}
\bar{J}(t)&=&-\frac{1}{16\pi^2}\int_0^1 dx~
\log\frac{M^2 - t x(1-x) - \Delta x}{M^2 - \Delta x}
\nonumber\\&=&
\frac{1}{32\pi^2}\left\{
2 + \frac{\Delta}{t}\log\frac{m^2}{M^2} -\frac{\Sigma}{\Delta}
\log\frac{m^2}{M^2} - \frac{\sqrt{\lambda}}{t}
\log\frac{(t+\sqrt{\lambda})^2 -
\Delta^2}{(t-\sqrt{\lambda})^2-\Delta^2}\right\}
{}~,
\nonumber\\
J^r(t) &=& \bar{J}(t) - 2k~,
\nonumber\\
M^r(t) &=& \frac{1}{12t}\left\{ t - 2 \Sigma \right\} \bar{J}(t)
+ \frac{\Delta^2}{3 t^2} \bar{J}(t)
+ \frac{1}{288\pi^2} -\frac{k}{6}
\nonumber\\&&
                               - \frac{1}{96\pi^2 t} \left\{
      \Sigma + 2 \frac{M^2 m^2}{\Delta}
     \log\frac{m^2}{M^2} \right\} ~,
\nonumber\\
L(t)&=& \frac{\Delta^2}{4t} \bar{J}(t)~,
\nonumber\\
K(t)&=&\frac{\Delta}{2t}\bar{J}(t)    ~,
\nonumber\\
H(t)&=&\frac{2}{3} \frac{L_9^r}{F^2} t + \frac{1}{F^2}[t M^r(t) - L(t)],
\nonumber\\
\Delta &=& M^2 - m^2~,
\nonumber\\
\Sigma &=& M^2 + m^2  ~,
\nonumber\\
\lambda&=&\lambda(t,M^2,m^2) ~=~ (t+\Delta)^2 - 4tM^2  ~.
\end{eqnarray}
In the text these are used with subscripts,
\begin{equation}
\bar{J}_{ij}(t)  =  \bar{J}(t)~~~\mbox{with}~~~M = M_i , m = M_j~
\end{equation}
and similarly for the other symbols.
The subtraction-point-dependent part
is contained in the constant $k$
\begin{equation}
k = \frac{1}{32\pi^2} \frac{M^2 \log \left( \frac{M^2}{\mu^2} \right)
                       - m^2 \log\left(\frac{m^2}{\mu^2}\right)}
   {M^2 - m^2},
\end{equation}
where $\mu$ is the subtraction scale.

In addition, in Section \ref{Kl3g} these functions and symbols
appear in a  summation
over loops $I$
with
\begin{eqnarray}
J_I(t) &=& \bar{J}(t) ~~~\mbox{with}~~~M = M_I , m = m_I       ~,
\nonumber\\
\Sigma_I&=&M_I^2 + m_I^2
\end{eqnarray}
and again similarly for the others.
There the combination $B_2$ appears  as well :
\begin{eqnarray}
B_2(t,M^2,m^2)&=&B_2 (t,m^2,M^2)
\\
&=&\frac{1}{288\pi^2} \left( 3\Sigma - t\right)
-\frac{\lambda(t,M^2,m^2) \bar{J}(t)}{12t}
+\frac{t\Sigma - 8M^2 m^2}{384\pi^2 \Delta} \log\frac{M^2}{m^2} ~.
\nonumber
\end{eqnarray}

The last formula to be defined is the three-propagator-loop integral
function $C(t_1,t_2,M^2, m^2)$, where one of the three external momenta
has zero mass and two of the propagators have the same mass $M$.
Here $t_1 = ( q_1 + q_2)^2$, $t_2 = q^2_2$ and $q_1^2 = 0$:
\begin{eqnarray}
C(t_1,t_2,M^2,m^2)&=& - i \int \frac{d^4p}{(2\pi)^d}
\frac{1}{(p^2 - M^2 ) ((p+q_1)^2 - M^2) ((p+q_1+q_2)^2 - m^2)}
\nonumber\\ &=&
-\frac{1}{16\pi^2}\int_0^1 dx \int_0^{1-x} dy
\frac{1}{M^2 - y ( \Delta + t_1 ) +  xy (t_1 - t_2) + y^2 t_1}
\nonumber\\
&=& \frac{1}{(4\pi)^2 (t_1 - t_2)}
\left\{ Li_2 \left( \frac{1}{y_+(t_2)}\right) +
  Li_2\left(\frac{1}{y_- (t_2)}\right)\right.
\nonumber\\& &  \left.
  - Li_2\left(\frac{1}{y_+(t_1)}\right)
  - Li_2\left(\frac{1}{y_-(t_2)}\right)\right\}~,
\nonumber\\
y_{\pm}(t)&=&\frac{1}{2t}\left\{ t + \Delta \pm
\sqrt{\lambda(t,M^2,m^2)}\right\}\;,
\end{eqnarray}
where $Li_2$ is the dilogarithm
\begin{equation}
Li_2(x) = - \int_0^1    \frac{dy}{y}\log(1-xy)   ~.
\end{equation}
\vspace{1cm}

\setcounter{equation}{0}
\addtocounter{zahler}{1}

\section{Decomposition of the hadronic tensors $I^{\mu \nu}$}
\label{kl2g}
Here we consider the tensors
\beq
I^{\mu \nu} =
\int dx e^{iqx+iWy} \left\langle
0 \mid T V^\mu_{em} (x) I^\nu_{4-i5}(y) \mid K^+(p)\right\rangle
\; \;, \; \; I=V,A \; \;
\eeq
and detail its connection with the matrix element (\ref{k3}).

The general decomposition of $A^{\mu\nu}, V^{\mu\nu}$
 in terms of Lorentz-invariant amplitudes reads \cite{BARDIN}, for $q^2
\neq 0$:
\beqa
\frac{1}{\sqrt{2}} A^{\mu\nu} &=& - F_K \left \{ \frac{(2W^\mu + q^\mu)
W^\nu}{M_K^2 - W^2} + g^{\mu\nu} \right \}
\nonumber \\
&&+ A_1 (q W g^{\mu\nu} - W^\mu q^\nu) + A_2 (q^2 g^{\mu\nu} - q^\mu
q^\nu)
\nonumber \\
&&+ \left \{ \frac{2 F_K (F_V^K(q^2)-1)}{(M_K^2-W^2)q^2}
+ A_3 \right\}
 (qWq^\mu - q^2 W^\mu) W^\nu
\label{A6}
\eeqa
and
\beq
\frac{1}{\sqrt{2}} V^{\mu \nu} = iV_1 \epsilon^{\mu \nu \alpha \beta}
q_\alpha p_\beta \;,
\label{A7}
\eeq
where the form factors $A_i(q^2,W^2)$ and $V_1(q^2,W^2)$ are analytic functions
of $q^2$ and $W^2$; $F^K_V (q^2)$ denotes the electromagnetic form factor of
the kaon $(F_V^K (0) = 1)$. The amplitudes $I^{\mu \nu}$  satisfy
the Ward identities (cf. Appendix \ref{ward})
\beqa
q_\mu A^{\mu \nu} &=& - \sqrt{2} F_K p^\nu \no \\
q_\mu V^{\mu \nu} &=& 0 \; \; .
\label{A8}
\eeqa
The decomposition in
Eq.~(\ref{A6}) may be derived as follows \cite{BARDIN}. One
first isolates the contribution from Bremsstrahlung off the kaon,
\beqa
A^{\mu \nu} &=& A_{\mbox{\tiny{IB}}}^{\mu \nu} + {\bar{A}}^{\mu \nu} \no \\
A_{\mbox{\tiny{IB}}}^{\mu \nu} &\doteq& -\sqrt{2} F_K \frac{F_V^K(q^2)}{M_K^2
-W^2} (2 W^\mu +q^\mu) W^\nu     \\
q_\mu A_{\mbox{\tiny{IB}}}^{\mu \nu}&=& -\sqrt{2}F_K F_V^K(q^2) W^\nu \; .
\no
\label{eq:decomp}
\eeqa
The general structure of the remainder ${\bar{A}}^{\mu \nu}$ is
\beqa
-{\bar{A}}^{\mu \nu}= A_1 W^\mu q^\nu +A_2 q^\mu q^\nu +A_3 q^2 W^\mu W^\nu
+A_4 g^{\mu \nu} +A_5 q^\mu W^\nu .
\label{eq:abar}
\eeqa
The factor $q^2$,
 which multiplies $A_3$, is dictated by the Ward identity Eq.
(\ref{A8}) which implies
\beqa
qW A_1 +q^2 A_2 +A_4 &=&\sqrt{2} F_K \no \\
q^2 qW A_3 +q^2 A_5  &=& \sqrt{2} F_K \left[1-F_V^K(q^2)\right] \; .
\eeqa
Eliminating $A_4$ and $A_5$ gives (\ref{A6}).

 In the
process (\ref{k1}) the photon is real. As a consequence,
only the two form factors $A_1(0,W^2)$ and $ V_1(0,W^2)$ contribute. We set
\beqa
A(W^2)& = &A_1(0,W^2)
\nonumber \\
V(W^2) &=& V_1 (0, W^2)
\label{A10}
\eeqa
and obtain, for the matrix element (\ref{k3}):
\beq
T= -iG_F/\sqrt{2} e {V_{us}}^\star \epsilon^\star_\mu \left \{ \sqrt{2} F_K
l_1^\mu - (V^{\mu\nu} - A^{\mu\nu}) l_\nu \right \}_{ \mid_{q^2=0}} \; \;
, \label{A2}
\eeq
with
\beqa
l^{\mu} & =& \bar{u} (p_\nu)\gamma^\mu  (1-\gamma_5) v(p_l)
\nonumber \\
l_1^\mu &=& l^\mu + m_l \bar{u} (p_\nu) (1+\gamma_5) \frac{2p_l^\mu + \not
\!{q} \gamma^\mu}{m_l^2 - (p_l + q)^2} v(p_l) \; \; .
\label{A3}
\eeqa
Grouping terms into an IB and a SD piece gives  (\ref{k3}), (\ref{k4}).
As a consequence of (\ref{A8}), $T$  is
 invariant
under the gauge transformation $\epsilon_\mu \rightarrow \epsilon_\mu +
q_\mu$.

The amplitudes $A_1, A_2$ and $V_1$ are related to the
corresponding quantities $F_A,R$ and $F_V$ used by the PDG \cite{PDG} by
\beq
- \sqrt{2} M_K (A_1, A_2, V_1) = (F_A, R, F_V).
\label{A9}
\eeq
The last term in (\ref{A6}) is omitted in \cite{PDG}. It
contributes to
processes with a virtual photon, $K^\pm \rightarrow l^\pm \nu_l l'^+ l'^-$.

Finally, the relation to  the notation used in
\cite{ke22,km21} is
\beqa
2 (A \pm V)^2 &=& (a_k \pm v_k)^2 \; \; \; \cite{ke22} \nonumber \\
\sqrt{2} (A,V)& =& (F_A,F_V) \; \; \; \cite{km21} \; \; .
\label{A11}
\eeqa

\vspace{1cm}
\setcounter{equation}{0}
\addtocounter{zahler}{1}

\section{ Ward identities}
\label{ward}

Here  we derive the Ward identity
\beqa
A^{\mu \nu}(q,p) &=&\int{dx e^{iqx +iWy}\langle 0|TV_{em}^\mu(x)
A_{4-i5}^\nu(y)|K^+(p)\rangle } \label{c1} \\
q_\mu A^{\mu \nu}&=&-\sqrt{2} F_Kp^\nu \; \; ; \; W \doteq p-q \label{c2}
\eeqa
without using formal manipulations with $T$-products and with
current commutators. Our treatment is similar to the considerations in
Ref. \cite{bell} and parallels the ones in Ref. \cite{haleut}.

The reason we wish to discuss this issue is the following.
Owing to the short-distance singularities of the
operator product of two currents, integrals
such as the one in Eq. (\ref{c1}) are {\it a priori} not well defined.
 As a result of
this, the Ward identities for $A^{\mu \nu}$ depend on the convention chosen for
$\langle 0|TV^\mu(x)A^\nu(y)|K \rangle .$ In CHPT, the Green functions are
 obtained through
(functional derivatives of) the generating functional $Z(v,a,s,p).$ This
functional is constructed such that it is invariant under transformations
generated by the vector currents (see Section \ref{CHPT}),
\beq
Z(v',a',s',p') = Z(v,a,s,p)\;,
\label{c3}
\eeq
where
\beqa
v_\mu' \pm a_\mu' &=& g(v_\mu \pm a_\mu)g^+ +ig\partial_\mu g^+ \nonumber \\
s'+ip'&=& g(s+ip)g^+ \nonumber \\
g(x)&\in&SU(3) \label{c4} \; \; .
\eeqa
The relation (\ref{c3}) embodies all the Ward identities associated with the
transformations (\ref{c4}). In particular, seagulls and Schwinger terms
 are automatically taken into account \cite{bell,haleut}.
For an infinitesimal transformation
\beqa
g&=&1+i\alpha +O(\alpha^2) \; \; , \nonumber \\
\delta v_\mu &=& \partial_\mu \alpha +i[\alpha,v_\mu] \doteq D_\mu \alpha \; \;
, \nonumber \\
\delta I &=& i[\alpha,I] \; \; ; \; I=a_\mu,s,p \;,
\eeqa
one obtains from (\ref{c3})
\beq
\left\langle
\alpha D_\mu \frac{\delta Z}{\delta v_\mu (x)}\right\rangle =
i\sum_{I=a_\mu,s,p} \left\langle
[\alpha,I(x)] \frac{\delta Z}{\delta I(x)}\right\rangle \;,
\label{c6}
\eeq
where
\beq
I=\frac{\lambda^a}{2} I^a \; \; , \; \; \frac{\delta Z}{\delta I} \doteq
\frac{\lambda ^a}{2}\frac{\delta Z}{\delta I^a}\;.
\eeq
The Ward identities for current correlation functions  are
obtained
from this general formula  by taking additional functional derivatives,
 evaluated at
\beq
v_\mu=a_\mu=p=0 \;, \; s={\mbox{diag}}(m_u,m_d,m_s) \; \; .
\label{c7}
\eeq
In the following we consider the isospin symmetry limit $m_u = m_d$. In order
to derive Eq.~(\ref{c2}), we
 consider the quantity
\beq
\Gamma^{\mu \nu \rho}(x,y,z) =
\langle 0|TV_{em}^\mu(x)A_{4-i5}^\nu(y)A_{4+i5}^\rho(z)|0\rangle \;,
\label{c8}
\eeq
which provides an off-shell extension of $A^{\mu \nu},$ and translate the Ward
identity for  $\Gamma^{\mu \nu \rho}$ into the corresponding relation for
$A^{\mu \nu}.$
 Differentiating (\ref{c6}) with respect to $a_\nu^{4-i5}(y)$ and
$a_\rho^{4+i5}(z)$ at (\ref{c7}), one obtains with $\alpha$ = $\frac{1}{3}
{\mbox{diag}}(2,-1,-1)$
\beq
\partial_\mu^x\Gamma^{\mu \nu \rho}(x,y,z) = -[\delta^4(x-y)
-\delta^4(x-z)]\langle 0|TA_{4-i5}^\nu(y)A_{4+i5}^\rho(z)|0\rangle  \; \; .
\label{c9}
\eeq
Next we introduce the Fourier transforms
\beqa
i\int{
dz e^{-ip^\star z} \langle 0|TA_{4-i5}^\nu (z)A_{4+i5}^\rho (0)|0\rangle
     }
&=& \frac{R_2^{\nu
\rho}(p^\star)}{M_{K}^2-{p^\star}^2}
\nonumber \\
\int{
dx dz e^{iqx+iW^\star y -i p^\star z} \Gamma^{\mu \nu \rho} (x,y,z)
     }
&=&
\frac{R_3^{\mu \nu \rho}(q,W^\star)}{M_{K}^2-{p^\star}^2}
\label{c10}
\eeqa
where
\beq
W^\star \doteq p^\star-q\; , \; {p^\star}^2 \neq M_{K}^2 \; .
\label{c11}
\eeq
By use of
\footnote{Again
note that we work with the Condon--Shortley--de Swart phase
convention.}
\beq
\lim_{p^\star \rightarrow p} \left\{\begin{array}{ll}
R_2^{\nu \rho}=2 F_K^2 p^\nu p^\rho \\
    R_3^{\mu \nu \rho} = \sqrt{2} F_K A^{\mu \nu}p^\rho
                                     \end{array} \right.
\label{c12}
\eeq
one finds from (\ref{c9})
\beq
q_\mu A^{\mu \nu}  = -\sqrt{2} F_K p^\nu \; .
\eeq

 The other Ward identities considered in this article may be
derived in an analogous manner.


\newpage
\begin{thebibliography}{99}
\newcommand{\PL}[3]{{Phys. Lett.}        {#1} {(19#2)} {#3}}
\newcommand{\PRL}[3]{{Phys. Rev. Lett.} {#1} {(19#2)} {#3}}
\newcommand{\PR}[3]{{Phys. Rev.}        {#1} {(19#2)} {#3}}
\newcommand{\NP}[3]{{Nucl. Phys.}        {#1} {(19#2)} {#3}}

\bibitem{DAFHB} The DAFNE Physics Handbook, Eds. L. Maiani, G. Pancheri
and N. Paver, INFN-Frascati (to appear).

\bibitem{Wein79}
 S. Weinberg, Physica 96A (1979) 327.

\bibitem{GLAP}
 J. Gasser and H. Leutwyler, Ann. Phys. 158 (1984) 142.

\bibitem{GLNP1}
 J. Gasser and H. Leutwyler, Nucl. Phys. B250 (1985) 465.

\bibitem{ABJB}  S.L. Adler, Phys. Rev. 177 (1969) 2426;\\
J.S. Bell and R. Jackiw, Nuovo Cimento 60A (1969) 47;\\
W.A. Bardeen, Phys. Rev. 184 (1969) 1848.

\bibitem{WZW}  J. Wess and B. Zumino, Phys. Lett. 37B (1971) 95;\\
E. Witten, Nucl. Phys. B223 (1983) 422.

\bibitem{DHH}
J.F. Donoghue and B.R. Holstein, \PR{D40}{89}{3700}.

\bibitem{Hol} B.R. Holstein, Phys. Rev. D41 (1989) 2829.

\bibitem{GLNP2}
 J. Gasser and H. Leutwyler, Nucl. Phys. B250 (1985) 517.

\bibitem{PDG}
 Review of Particle Properties, Phys. Lett. B239 (1990).

\bibitem{ke22}
J. Heintze et al., \NP{B149}{79}{365}.

\bibitem{km21}
Y. Akiba et al., \PR{D32}{85}{2911}.

\bibitem{brym}
S.G. Brown and S.A. Bludman, \PR{B136}{64}{1160};\\
D.A. Bryman et al., Phys. Rep. C88 (1982) 151.

\bibitem{BEGRep}
J. Bijnens, G. Ecker and J. Gasser, Semileptonic kaon decays, in Ref.
\cite{DAFHB}.

\bibitem{ke21}
K.S. Heard et al., \PL{55B}{75}{324}.

\bibitem{km22}
V.V. Barmin et al., Sov. J. Nucl. Phys. 47 (1988) 643.

\bibitem{BARDIN}
D.Yu. Bardin and E.A. Ivanov, Sov. J. Part. \& Nucl. 7 (1976) 286.

\bibitem{EGPR}
G. Ecker, J. Gasser, A. Pich and E. de Rafael, \NP{B321}{89}{311};\\
G. Ecker, J. Gasser, H. Leutwyler, A. Pich and E. de Rafael,
\PL{B223}{89}{425};\\
J.F. Donoghue, C. Ramirez and G. Valencia, \PR{D39}{89}{1947};\\
M. Praszalowicz and G. Valencia, \NP{B341}{90}{27}.

\bibitem{Low}
F.E. Low, Phys. Rev. 110 (1958) 974.

\bibitem{KRISHNA} S. Krishna and H.S. Mani, Phys. Rev. D5 (1972) 678.

\bibitem{BCC}
J. Bijnens and F. Cornet, Nucl. Phys. B296 (1986) 557.

\bibitem{DIAMANT} A.M. Diamant-Berger et al., Phys. Lett. 62B (1976) 485.

\bibitem{scheck} A.~Kersch and F.~Scheck, \NP{B263}{86}{475}.

\bibitem{ATIYA} M.S. Atiya et al., Phys. Rev. Lett. 63 (1989) 2177.

\bibitem{egli}
S. Egli et al., Phys. Lett. B175 (1986) 97 and B222 (1989) 533.

\bibitem{Thesis} S. Egli, Der seltene Pion-Zerfall $\pi^+ \to
e^+ e^+ e^- \nu_e$, Thesis, Univ. Z\"urich, 1987.

\bibitem{Doncel}
M.G. Doncel, Phys. Lett. 32B (1970) 623.

\bibitem{FFS} E. Fischbach and J. Smith, Phys. Rev. 184 (1969)
 1645; \\
 H.W. Fearing, E. Fischbach and J. Smith, Phys. Rev. D2 (1970) 542.

\bibitem{Franz}
P. Franzini, in Ref. \cite{DAFHB}.

\bibitem{Bijnens}
J. Bijnens, Nucl. Phys. B337 (1990) 635.

\bibitem{Rigg}
C. Riggenbach,  J. Gasser, J.F. Donoghue and B.R. Holstein, Phys. Rev.
D43  (1991) 127.

\bibitem{coga}
J. Bijnens, G. Colangelo and J. Gasser, in preparation.

\bibitem{lroos}
H. Leutwyler and M. Roos, Z. Phys. {C25} (1984) 91; \\
J.F. Donoghue, B.R. Holstein and S.W. Klimt, \PR{D35}{87}{934}.

\bibitem{bell}
J.S. Bell, Nuovo Cimento 50A (1967) 129.

\bibitem{haleut}
P. Hasenfratz and H. Leutwyler, Nucl. Phys. B343 (1990) 241.

\end{thebibliography}
\end{document}